 \documentclass[final,5p,times,twocolumn,authoryear]{elsarticle}

\usepackage{amssymb}
\usepackage{amsmath}
\usepackage{aas_macros}
\usepackage{bbm}
\usepackage{bm}
\usepackage{multirow}
\usepackage{booktabs}
\usepackage{xurl}
\usepackage{hyperref}
\hypersetup{
    colorlinks=true,
    linkcolor=blue,
    citecolor=blue,
    urlcolor=blue
}
\usepackage{xcolor} 

\journal{Astronomy $\&$ Computing}

\begin{document}

\begin{frontmatter}



\title{Multi-Scale Contrastive Attention for Light-Curve Representation Learning}

\author[inst1]{Torsha Majumder}
\affiliation[inst1]{organization={Independent Researcher},
            country={India}}
\author[inst2]{Konstantin Malanchev}
\affiliation[inst2]{organization={McWilliams Center for Cosmology and Astrophysics, Carnegie Mellon University},
             addressline={5000 Forbes Avenue},
             city={ Pittsburgh},
             postcode={PA 15213},
             state={Pennsylvania},
             country={USA}}

\author[inst3]{Emille E. O. Ishida}
\affiliation[inst3]{organization={Université Clermont Auvergne, CNRS/IN2P3, LPCA},
            city={Clermont-Ferrand},
            postcode={F-63000}, 
            country={France}}

\begin{abstract}
Current and next-generation time-domain surveys demand automated techniques capable of analyzing millions of light curves, observed in multiple filters, without relying on exhaustive human annotation or scarce spectroscopic follow-up. We present \texttt{Astra-CLR}, an attention-based, self-supervised contrastive learning framework which enables the representation of raw light curves into a highly discriminative  latent space. Pre-trained on $\sim$2.1 million unlabeled Zwicky Transient Facility light curves, the framework utilizes partial light curves as input sequences to generate asymmetric, multi-scale temporal views (explicitly contrasting shorter sequences against longer ones) forcing the network to learn a robust ``local-to-global'' mapping strategy. Furthermore, we introduce a novel multi-view late fusion architecture that extends the model to efficiently handle longer light curves with larger numbers of observations while accommodating the different cadences associated with each filter. The discriminatory power of the resulting representations was evaluated by using them as input to a Multinomial Logistic Regression classifier, trained to identify 12 broad classes of variability. Final accuracy achieved $\sim 0.70$. When applying a label-efficient, partial top-layer fine-tuning strategy, the topological structure of the latent space is significantly refined, boosting results to $\sim$0.77. \texttt{Astra-CLR} is the first publicly available multi-filter time-series Transformer trained exclusively on real ZTF light curves. Results presented here demonstrate that it provides an ideal foundation for the development of end-to-end pipelines, taking into account color evolution and respecting the inhomogeneous nature of astronomical light curve sampling. 
\end{abstract}



\begin{keyword}
Self-supervised learning \sep Representation learning \sep Time series analysis \sep Variable stars



\end{keyword}

\end{frontmatter}




\begin{figure}[t!]
	\centering \includegraphics[width=0.95\columnwidth,trim=0.5cm 1cm 0.5cm 1cm, clip]{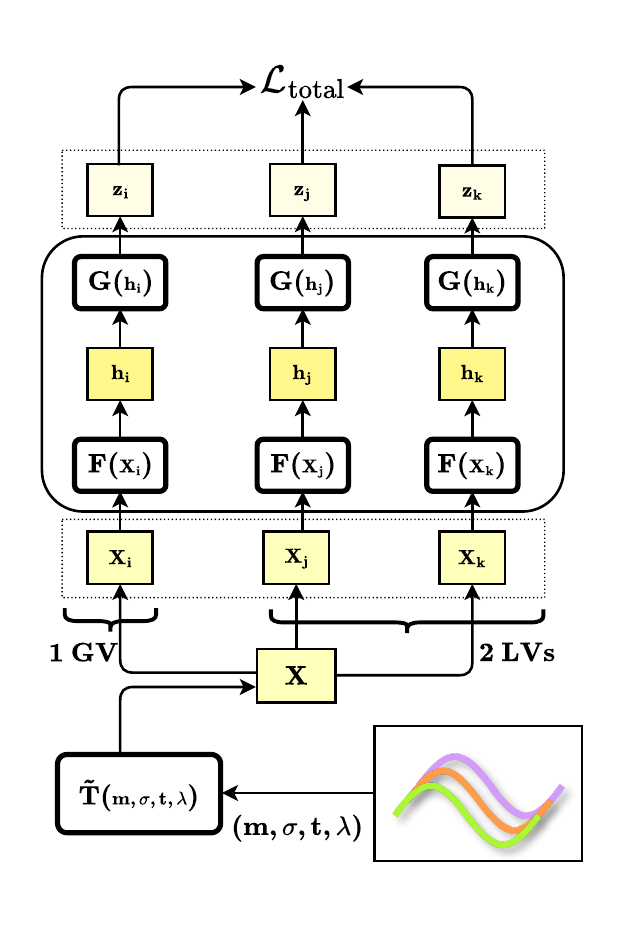}	
	\caption{High-level schematic of the \texttt{Astra-CLR} contrastive learning framework. A raw light curve, parameterized by $(m, \sigma, t, \lambda)$, is processed through the input representation pipeline, $\mathbf{\tilde{T}(\cdot)}$, to generate multi-scale temporal views (detailed in Figure \ref{fig:inpu-emb}). These asymmetric views---comprising one Global View ($\mathbf{GV, X}_i$) and two Local Views ($\mathbf{LVs, X}_j, \mathbf{X}_k$)---collectively form a \textit{positive group}. The augmented views are simultaneously passed through the shared \texttt{AstraNet} encoder, which consists of a backbone $\mathbf{F}(\cdot)$ and a projection head $\mathbf{G}(\cdot)$. The backbone extracts a fixed-length global representation ($\mathbf{h} \in \mathbb{R}^{d_{\text{model}}}$) for each view, which is then mapped by the projection head to a lower-dimensional latent space ($\mathbf{z} \in \mathbb{R}^{d_{\text{proj}}}$; see Figure \ref{fig:astra-net}). Finally, the network is optimized via the NT-Xent objective ($\mathcal{L}_{\text{total}}$), which maximizes the latent similarity among all projections $\mathbf{z} \in \{\mathbf{z}_i, \mathbf{z}_j, \mathbf{z}_k\}$ within the same \textit{positive group} as detailed in Section \ref{sec:nt-xent-loss}.}
	\label{fig:astra-arch}
\end{figure}

\section{Introduction}
\label{sec:introduction}
In the current era of time-domain astronomy, active wide-field surveys like the Zwicky Transient Facility (ZTF; \citealt{Bellm_ZTF_2019}) and the Gaia mission \citep{prusti2016gaia} are already generating enormous archives of unlabeled photometry. This data deluge will only accelerate into the petabyte scale with next-generation observatories such as the Vera C. Rubin Observatory's Legacy Survey of Space and Time (LSST; \citealt{lsst}) and the Nancy Grace Roman Space Telescope \citep{rose2021roman}. However, categorizing and analyzing this unprecedented volume of unlabeled photometric time-series data remains a formidable bottleneck. While spectroscopic follow-up provides definitive physical insights into these objects, acquiring spectra is observationally expensive, highly competitive, and strictly limited in volume. Consequently, to fully harness the scientific potential of modern time-domain surveys, the astronomical community requires algorithms capable of disentangling complex, discriminative patterns strictly from raw photometric light curves.

In response to this bottleneck, recent efforts have shifted toward Self-Supervised Learning (SSL). By leveraging the attention mechanisms of the Transformer architecture \citep{attention}, researchers have begun training foundational representation models on millions of unannotated light curves. While these models effectively generate latent embeddings for downstream tasks, current astronomical SSL literature exhibits distinct methodological constraints. Cross-modal frameworks, such as $\mathrm{AstroM}^3$ \citep{astrom3}, MAVEN \citep{maven}, and AstroCLIP \citep{astroclip}, effectively learn from the joint distribution of photometry and spectra (and occasionally metadata), but inherently experience a performance dip when only raw photometry is available. Supervised approaches, such as ATAT \citep{atat}, similarly rely on auxiliary metadata (e.g., redshift, host galaxy properties, Milky Way extinction) alongside light curves to guide training. Conversely,  existing frameworks, including FALCO \citep{zuo2026}, ASTROMER \citep{astromer}, ASTROMER 2 \citep{astromer2}, ASTRAFier \citep{gregory2026} and StarCLR \citep{ding2026}, operate exclusively on photometry but are limited to single-filter light curves. StarEmbed \citep{li2025} compares results using domain-specific time-series models against generic ones, in a single-filter analysis. \citet{allam2024} proposes \texttt{t2}, where a multi-dimensional Gaussian Process interpolation and 1D convolution embedding precede the Transformer network, thus translating the multi-filter light curve to a uni-dimensional input layer.
While the ATCAT \citep{atcat} framework is designed to process multiple photometric filters simultaneously, it still achieves its highest performance by integrating auxiliary tabular data alongside the light curves. 
Thus, the problem of developing a fully unsupervised representation learning model for multi-filter light curve data remains an open challenge.

We address this challenge with \texttt{Astra-CLR} (\textbf{A}ttention-based \textbf{S}elf-supervised \textbf{T}ime-series \textbf{R}epresentation \textbf{A}rchitecture with \textbf{C}ontrastive \textbf{L}ea\textbf{r}ning), a framework designed to operate natively on multi-filter light curves. Pre-trained on $\sim 2.1$ million unlabeled light curves from the Zwicky Transient Facility \citep[ZTF;][]{Bellm_ZTF_2019}, Zubercal Data Release 16, our framework requires no spectral or metadata intervention to untangle complex astronomical phenomena. The primary engine driving \texttt{Astra-CLR} is a suite of domain-specific data augmentation strategies. We leverage these augmentations to introduce a novel \textit{multi-scale temporal view} mechanism \citep[analogous to the \textit{multi-crop} paradigm in computer vision,][]{swav} which forces the attention-based \texttt{AstraNet} encoder to execute a strict \textit{local-to-global} mapping strategy. To further overcome the extreme cadence variability of surveys, we introduce a \textit{multi-view late fusion} architecture capable of mapping highly variable observational data into fixed-length, information-dense representations. Ultimately, \texttt{Astra-CLR} generates an exceptionally robust latent space that enables high-performance classification using only simple linear models. 
A high-level schematic of the \texttt{Astra-CLR} framework is presented in Figure \ref{fig:astra-arch}, with its core mechanisms and architectural components detailed in the subsequent sections.

The remainder of this paper is structured as follows. Section \ref{sec:data} details the acquisition of the Zubercal photometry, our cross-matching procedures, and dataset serialization. Section~\ref{sec:methods} describes our domain-specific data augmentations, the core \texttt{Astra-CLR} contrastive framework, and the multi-view late fusion feature extraction strategy. Section~\ref{sec:training-config} outlines the hyperparameter configurations and our distributed training strategy. In Section~\ref{sec:results}, we present the downstream evaluation of the pre-trained embeddings alongside our label-efficient fine-tuning results. Section \ref{sec:discussion} provides a critical interpretation of our key findings, while Section~\ref{sec:conclusion} summarizes our conclusions and outlines avenues for future work. Section~\ref{sec:software} notes the availability of the source code, inference scripts, and model weights to ensure full reproducibility. Finally, \ref{sec:ablation_studies} provides comprehensive ablation studies empirically validating our core architectural design choices.

\begin{table*}[t!]
    \centering
    \renewcommand{\arraystretch}{1.3} 
    \caption{Class distribution of the complete, pre-processed Gaia-Zubercal cross-matched dataset. The hierarchical classification of the variable stars—encompassing the broad variability type and specific variability group—follows the conventions established by \citet{eyer2023gaiadr3} and \citet{Rimoldini_GaiaVariability_2023}, while the designated VSX classification nomenclature is adopted from the International Variable Star Index \citep{watson2006international, samus2017general}. The last column denotes the total count of pre-processed light curves retained per class prior to the pre-training and validation split.}
    \label{tab:variable_dist}
    \vspace{5pt}
    \begin{tabular*}{\textwidth}{@{\extracolsep{\fill}}lccr}
        \toprule
        \textbf{Variability Type} & \textbf{Group} & \textbf{Classification Name} & \textbf{Count} \\
        \midrule \midrule 
        \vspace{1pt}
        Active galactic nuclei or QSO & AGN & AGN & 391 080 \\
        \vspace{1pt}
        Eclipsing binaries & Eclipsing systems & ECL & 531 584 \\
        \vspace{8pt}
        
        Cataclysmic variables & Eruptive/Cataclysmic & CV & 1 703 \\

        Solar-like variability & Rotation & SOLAR\_LIKE & 791 096 \\
        RS Canum Venaticorum & Rotation & RS & 270 340 \\
        \vspace{8pt}
        Ellispoidal variations & Rotation & ELL & 5 236 \\
        
        $\delta$ Scuti/$\gamma$ Doradus/SX Phoenicis stars & Pulsation  & DSCT/GDOR/SXPHE & 281 335 \\
        Long-period variables & Pulsation  & LPV & 89 258 \\
        RR Lyrae stars & Pulsation  & RR & 40 201 \\
        \vspace{8pt}
        Cepheids & Pulsation  & CEP & 1 534 \\

        Short-timescales & Other & S & 160 669 \\
        Young stellar objects & Other & YSO & 29 531 \\
        
        \bottomrule
    \end{tabular*}
\end{table*}

\section{Data}
\label{sec:data}
The efficacy of any robust SSL framework depends fundamentally on the scale, diversity, and quality of its pre-training data. In this section, we detail the end-to-end data acquisition and engineering pipeline constructed for the \texttt{Astra-CLR} framework. We first outline the extraction of multi-filter time-series photometry from the Zubercal data release and our rigorous cross-matching methodology against the Gaia DR3 catalog (Section \ref{sec:zubercal-cross-matching}). Subsequently, we detail the preprocessing and serialization strategies employed to transform the raw observations into a highly optimized format, followed by a dynamic standardization step applied during data loading, immediately prior to the augmentation pipeline (Section \ref{sec:data-prep}).
\subsection{The Zubercal Photometry and Object Selection}
\label{sec:zubercal-cross-matching}
To construct the photometric time-series dataset for the \texttt{Astra-CLR} framework, we utilize the Zwicky Transient Facility (\textsc{ZTF}; \citealt{Bellm_ZTF_2019}) Zubercal Data Release 16\footnote{\url{http://atua.caltech.edu/ZTF/Zubercal.html}} (DR16), which provides multi-filter light curves for over a billion objects.
While Zubercal DR16 spans the same observational baseline as the standard ZTF DR16 release, its underlying photometry is generated via a rigorous global recalibration process, resulting in several advantages over the standard ZTF pipeline \citep{masci2019zwicky}.
First, whereas ZTF photometry is calibrated on a localized, field-by-field basis, which often leads to zero-point inconsistencies, Zubercal performs a global ``ubercalibration'' \citep{2008ApJ...674.1217P} using the Pan-STARRS1 (PS1) DR2 catalog. This reduces systematic differences in the photometry of the same object arising from overlapping observational fields.
Second, Zubercal does not require an object's prior inclusion in the standard ZTF reference catalog (though it still requires the object to be present in PS1 DR2), which recovers additional detections and produces more complete light curves.
Finally, individual detections are already cross-matched to the PS1 DR2 object catalog, making Zubercal catalogs easier to work with than the standard ZTF data releases.

Since the vast majority of ZTF light curves do not exhibit significant photometric variability, we filter the dataset by cross-matching sources with the Gaia DR3 variability catalog \citep{GaiaCollaboration_GaiaDR3_2023, Rimoldini_GaiaVariability_2023}. As the largest homogeneous all-sky catalog of variable stars, Gaia DR3 serves as an ideal reference for isolating variable sources in ZTF. Furthermore, this cross-matching resulted in robust, machine-learning-derived labels. These labels serve as the ground truth for all the downstream tasks of \texttt{Astra-CLR}, encompassing label-efficient partial fine-tuning and the evaluation of both the pre-trained and fine-tuned \texttt{Astra-CLR} representations (see Section \ref{sec:results}).

To facilitate efficient large-scale cross-matching, we access the PS1 and Zubercal data via publicly available HATS catalogs\footnote{\url{https://data.lsdb.io}}, whereas the Gaia DR3 variability catalog (I/358/vclassre) is retrieved via VizieR \citep{Ochsenbein_VizieR_2000}. Utilizing the LSDB framework \citep{Caplar_LSDB_2025}, we first filter the Zubercal photometry to retain only high-quality observations---specifically, those free of calibration errors (\texttt{info} $=0$) and those corresponding to good observational conditions (\texttt{flag} $=0$). The filtered Zubercal time-domain photometry is then joined with the PS1 DR2 reference catalog using their shared PS1 object identifiers. Finally, we perform a cross-match against the Gaia DR3 variability catalog, assigning each Gaia object to its nearest PS1 counterpart within a $1$~arcsecond search radius. This rigorous cross-matching pipeline successfully paired 4 975 726 sources out of the original 9 976 881 Gaia variable objects.

\begin{figure}[t!]
	\centering \includegraphics[width=0.95\columnwidth, trim=0.5cm 0.5cm 1.2cm 0.5cm, clip]{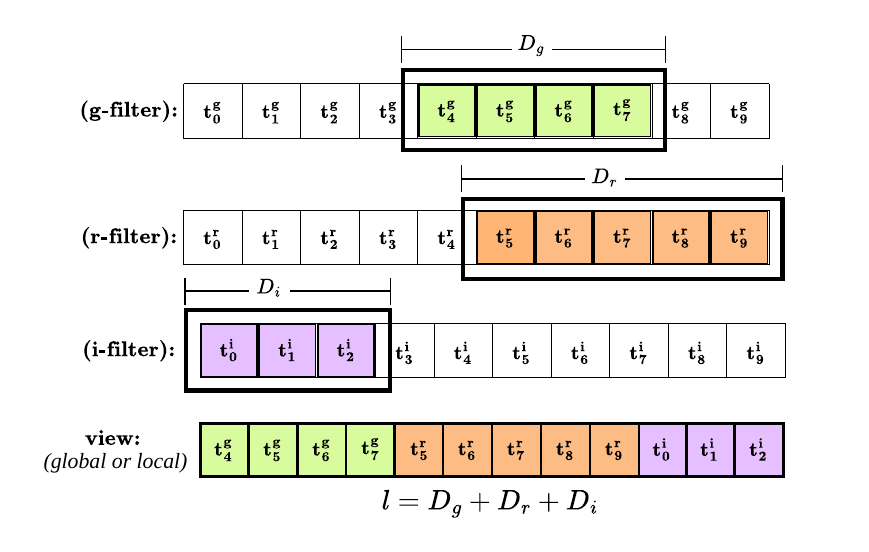}	
    	\caption{Schematic diagram of the Multi-Filter Random Window augmentation strategy used to generate a temporal view of fixed length $l=12$. For illustrative purposes, we assume the original light curve contains 10 consecutive detections per photometric filter ($g$, $r$, $i$). The algorithm extracts a random, continuous temporal window from each filter. As denoted by the colored blocks, the strategy samples windows [$t_4^g$--$t_7^g$], [$t_5^r$--$t_9^r$], and [$t_0^i$--$t_2^i$] of lengths $D_g$, $D_r$, and $D_i$, respectively. These filter-specific windows are subsequently concatenated in the strict order of $[g, r, i]$ to construct the final unified view, perfectly satisfying the target sequence length ($l = D_g + D_r + D_i = 12$).}
    \label{fig:random-win}
\end{figure}

\subsection{Data Serialization and Preprocessing}
\label{sec:data-prep}
To construct the final pre-training dataset and maximize the computational efficiency of the \texttt{Astra-CLR} data augmentation pipeline, we execute a secondary LSDB pipeline that applies rigorous preprocessing and selection criteria, serializing the filtered light curves into the TFRecord format to optimize data throughput. First, to make our augmentation strategies robust, we restrict our dataset to light curves containing a minimum of $200$ detections in both \textit{g} and \textit{r} filters of ZTF. We do not explicitly constrain the \textit{i}-filter due to its sparsity in the number of detections, which is inherited from its lower observing cadence. This $200$-detection threshold was specifically chosen to guarantee that each retained light curve contains sufficient temporal density to reliably generate distinct global and local views without relying on excessive zero-padding. Second, although SSL models are generally class-agnostic and do not strictly require a balanced dataset for pre-training, we restrict our pre-training dataset to objects that contain at least 1 000 light curves after applying the minimum detection threshold criteria. This selection strategy allows us to filter out any extreme outliers in the dataset and ensures that the model learns robust representations from statistically significant populations. Third, before generating the input embeddings for \texttt{AstraNet} encoder, we represent each light curve as a combination of its apparent magnitude ($m$), apparent magnitude error ($\sigma$), the time of observation in MJD ($t$), and the effective wavelength ($\lg\lambda$) of the corresponding ZTF filter. Since the \texttt{Astra-CLR} uses a multi-filter input strategy, incorporating the effective wavelength allows the model to naturally distinguish between filters, enabling it to implicitly derive color information. The detections in each light curve are then structured as a unified sequence, grouped first by filter, in the order of \textit{g}-filter, \textit{r}-filter, and \textit{i}-filter, and subsequently sorted in ascending observational time within each filter group. Fourth, to optimize the memory bandwidth and training throughput, we explicitly cast all input tensors to $32$-bit floating-point precision (\texttt{float32}). To prevent numerical precision loss during this downcasting, we subtract a constant offset of 58 000  from all MJD values.

Following these rigorous preprocessing steps, our total photometric time-series data comprises 2 593 567 light curves. While we strictly withheld the cross-matched Gaia DR3 labels during the self-supervised pre-training phase, we present the imbalanced distribution of the twelve distinct variable star classes in Table \ref{tab:variable_dist}. From this massive collection, we utilize 2 073 458 light curves for self-supervised pre-training, reserving the remainder for validation. Both the Gaia-Zubercal cross-matched HATS catalog and the fully processed TFRecord datasets are open-source and publicly available\footnote{\url{https://huggingface.co/datasets/snad-space/astra-zubercaldr16_gaiadr3vclassre}}.

During data loading and prior to data augmentation, we apply a standardization step. We dynamically center the apparent magnitude ($m$) of each light curve to ensure that the model also considers the relative magnitude fluctuations ($\sigma$) rather than the baseline magnitude. For a given light curve comprising $L$ total detections, we calculate the weighted mean magnitude using inverse-variance weighting. The weights are derived from the photometric uncertainties ($\sigma_j$) associated with each $j$-th apparent magnitude measurement ($m_j$) to reduce the influence of noisy data by assigning less weight to observations with higher variance. The centered apparent magnitude ($m'$) is then defined as the deviation from its weighted mean ($ \bar{m}_w$):
    
\begin{subequations} 
\label{eq:standardization_step}
    \begin{align}
        \bar{m}_w &= \frac{\sum_{j=1}^{L} \frac{m_j}{\sigma_{j}^2}}{\sum_{j=1}^{L} \frac{1}{\sigma_{j}^2}} \label{eq:weighted_mean_sub}, \\[1em]
        m' &= m - \bar{m}_w. \label{eq:centered_mag_sub}
    \end{align}
\end{subequations}

\begin{figure*}[t!]
	\centering 
	\includegraphics[width=\textwidth, trim=0.cm 0.cm 0.2cm 0.cm, clip]{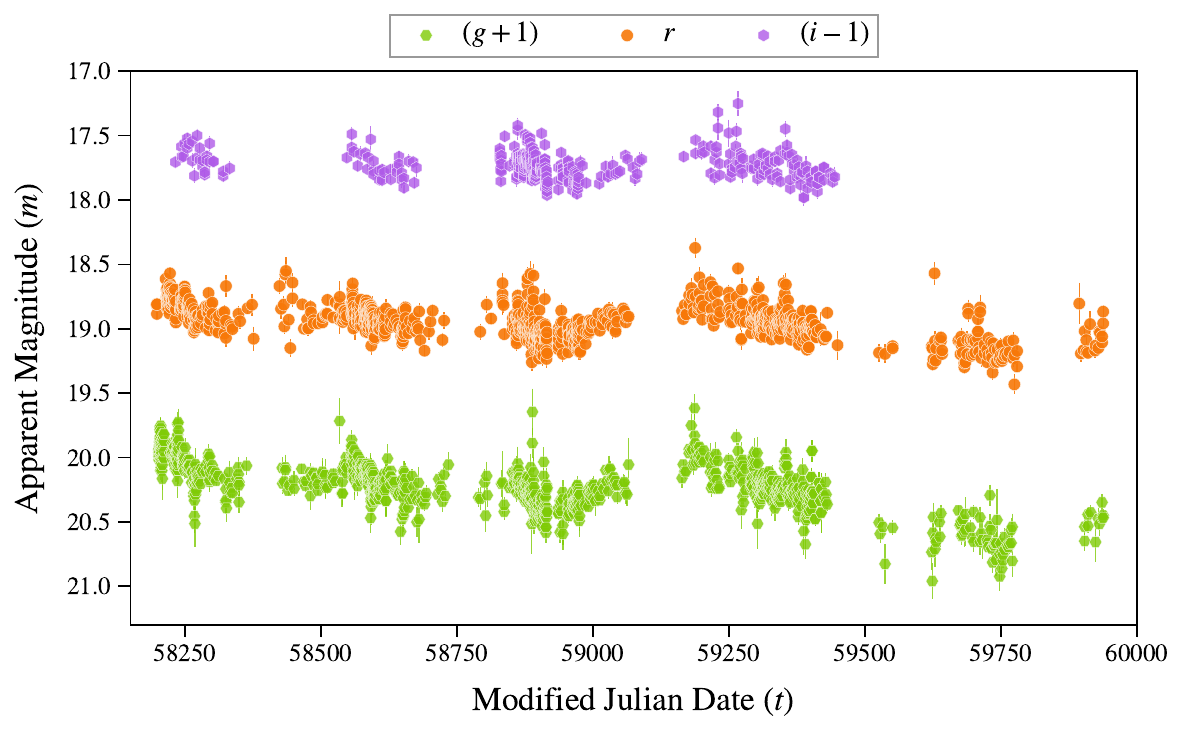}	
	\caption{The original, unaugmented Zubercal light curve of a source classified as an AGN in the Gaia DR3 variability catalog. The object corresponds to PS1 DR2 ID 174881967724736886 and Gaia DR3 ID 1564752621402023552, with equatorial coordinates ($\alpha = 196.77249669015^\circ$, $\delta = +55.73835489225^\circ$) derived from Gaia DR3. This raw, multi-filter light curve serves as the baseline reference prior to the application of the data preprocessing and augmentation pipeline.}
    \label{fig:original_lc}
\end{figure*}
\section{Methods} 
\label{sec:methods}

In this section, we present the complete architecture and training pipeline of our proposed framework. We begin by detailing the domain-specific data augmentation strategies, explicitly tailored for astronomical time-series, which are necessary to generate the robust, asymmetric, multi-scale temporal views that drive contrastive representation learning (Section \ref{sec:data-aug}). Following this, we introduce the core \texttt{Astra-CLR} framework (Section \ref{astra-clr-framework}), unpacking its primary components and operational mechanisms. This includes the formulation of the input embeddings (Section \ref{sec:astra-clr-emb}), and the \texttt{AstraNet} encoder used to extract contextual features from each sequence (Section \ref{sec:astra-net}). We then formalize the pre-training dynamics by detailing the NT-Xent contrastive loss function used to align representations in the latent space (Section \ref{sec:nt-xent-loss}), alongside the optimization mechanism and learning rate schedules required for stable model convergence (Section \ref{optandlr}). Finally, we present our multi-view late fusion strategy (Section \ref{late_fusion}), a novel feature extraction mechanism designed to map variable-length light curves into fixed-length representations while maximally capturing their long-term temporal dependencies for all downstream evaluations.

\begin{figure*}[t!]
	\centering 
	\includegraphics[width=\textwidth,trim=1.cm 1.cm 1.cm 0.8cm, clip]{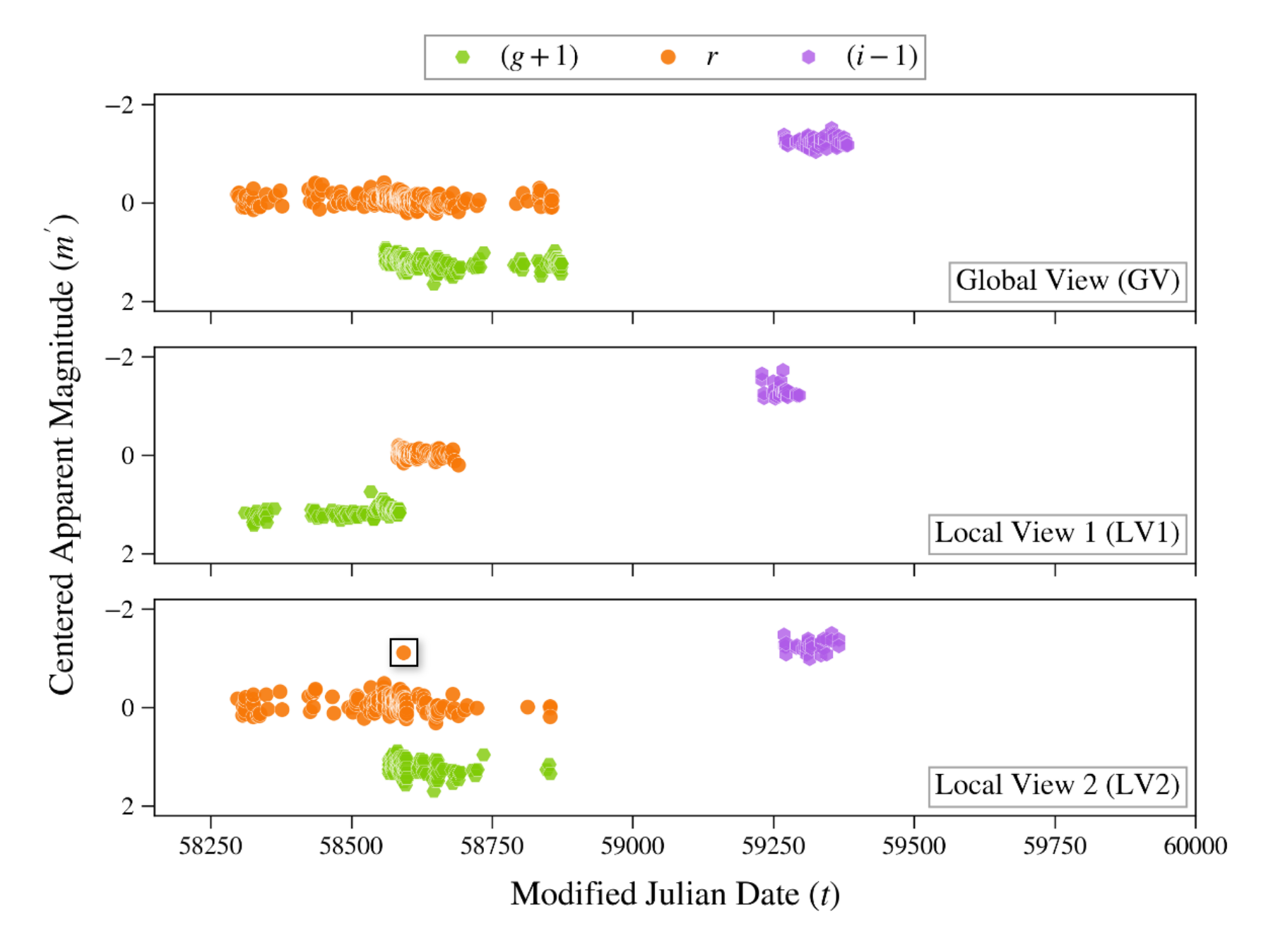}	
	\caption{Three augmented views generated from the baseline AGN light curve (Figure \ref{fig:original_lc}), collectively forming a single \textit{positive group} for contrastive pre-training. Following data preprocessing, the y-axis now reflects the centered apparent magnitude ($m'$). \textbf{Top:} A Global View (GV) of sequence length $l=700$ ($D_g=300, D_r=350, D_i=50$), augmented with Gaussian noise ($\sigma_{\text{noise}}=0.05$). \textbf{Middle:} A Local View (LV1) of length $l=350$ ($D_g=150, D_r=175, D_i=25$), representing a shorter, dense temporal sequence extracted via the multi-filter random window strategy. \textbf{Bottom:} A heavily augmented, sparse second Local View (LV2). While initially sampled from a wider sequence length of $l=700$ ($D_g=300, D_r=350, D_i=50$), the application of time-bin masking ($q=5$, $p=50\%$) explicitly forces the model to ignore half the temporal bins, reducing the effective unmasked sequence length to $l \approx 350$. This view is additionally subjected to higher Gaussian noise ($\sigma_{\text{noise}}=0.10$) and the synthetic injection of a photometric outlier (explicitly highlighted by the bounding box).}
	\label{fig:aug_lc}
\end{figure*}

\subsection{Data Augmentation for Contrastive Representation Learning}
\label{sec:data-aug}
State-of-the-art SSL frameworks in computer vision---such as SimCLR \citep{simclr}, MoCo \citep{mocov1,mocov2}, SwAV \citep{swav}, and DINO \citep{dino}---rely heavily on well-established geometric and pixel-level augmentation techniques, such as random cropping, rotation, color jittering, Gaussian blur, etc. For instance, AstroCLIP \citep{astroclip} is a Vision Transformer (ViT) adapted from \citet{vit}, pre-trained on galaxy images utilizing the DINOv2 \citep{dinov2} approach, where multiple augmented views are generated using these established strategies. However, designing analogous strategies for astronomical time series is a significant challenge due to the irregular sampling of the data, varying survey cadences, and the inherent noise of photometric light curves.

Despite these challenges, recent advances in astronomical representation learning have proposed various custom augmentation pipelines to generate augmented versions of the same data. To reduce overfitting, models such as ATCAT \citep{atcat} employ a stacked set of transformations applied at random rates. These include random subsampling, flux and time scaling, redshifting, and the addition of random noise. ATAT \citep{atat} introduces Masked Temporal Augmentation (MTA), which randomly truncates light curves at specific time cutoffs (e.g., $t \in \{8, 128, 2048\}$ days). MTA forces the model to learn from incomplete data, optimizing it specifically for early classification performance by simulating real-world alert scenarios where only a few days of data are available. Masked sequence modeling (MSM) frameworks like ASTROMER \citep{astromer} and ASTROMER 2 \citep{astromer2} adapt natural language processing (NLP) techniques by randomly replacing $50\%$ of the temporal positions with a standardized mix of masked tokens, true values, and random intra-sequence magnitudes. This probing technique effectively prevents the network from learning a direct identity mapping and improves its robustness against noise. Furthermore, frameworks such as MAVEN \citep{maven} and $\text{AstroM}^3$ \citep{astrom3} leverage a cross-modal contrastive learning objective. Conceptually, they are analogous to CLIP \citep{clip}, in that they employ distinct observational modalities---specifically time-series photometry, spectra, and available metadata belonging to the same object---to construct a shared, joint representation space. Even though these architectures utilize cross-modal alignment, they are still critically reliant on standard data augmentation techniques applied to the individual modalities to ensure representational robustness. For instance, to augment the data, MAVEN applies Gaussian noise to photometric and spectroscopic observations during each training iteration, with a standard deviation matching the reported observational errors. Finally, approaches like Multiband-NN \citep{multiband-nn} utilize full synthetic light curve generation as a mechanism for data augmentation to address the severe class imbalance in variable star populations.

Ultimately, the success of any SSL framework is grounded in its data augmentation strategies. The breakthrough of SimCLR proved that the careful composition of augmentations is critical for learning robust representations. Building on this, frameworks like SwAV and DINO introduced the multi-crop strategy, generating multiple local and global views of a single image. This approach drastically improved representation quality without exceeding compute budgets by forcing the model to learn a \textit{local-to-global} matching strategy. In \texttt{Astra-CLR}, we adapt this multi-crop paradigm to astronomical time-series by extracting multi-scale temporal views---multiple sequence views of variable lengths---from a single light curve. To generate these views from a single underlying sequence, we implement a stacked data augmentation pipeline that encourages the model to learn robust, time-invariant representations. These encompass both magnitude-level augmentations (analogous to pixel-level transformations) and temporal augmentations (analogous to geometric cropping). We define and detail these specific augmentation strategies below.

\begin{enumerate}
    \item \textit{Random Noise}: We perform magnitude-level augmentation by applying Gaussian noise to the centered apparent magnitudes ($m'$) during each training iteration to ensure that the model does not overfit to specific observational conditions. The injected noise is sampled from a normal distribution $\mathcal{N}(0, \sigma_{\text{noise}})$, where the standard deviation $\sigma_{noise}$ is a user-defined value within the range $[0.0, 0.2)$ configured to simulate the inherent unpredictability of atmospheric interference and instrumental noise.
    \item \textit{Photometric Outlier}: Real-world time-domain surveys frequently contain spurious detections resulting from image subtraction artifacts, saturated pixels, or cosmic rays \citep{sreejith2026dataset}. We force our architecture to remain invariant to such anomalies by randomly introducing simulated photometric outliers into the light curve, which serves as an additional magnitude-level augmentation. First, we define an empirical magnitude saturation limit based on the overall Zubercal data distribution used for pre-training. We then simulate an outlier by injecting a false detection into the light curve; we calculate its magnitude by drawing a random value from a uniform distribution and subtracting it from the saturation limit (e.g., $m'_{\text{sat}} - \mathcal{U}(0,0.5)$), ensuring the injected detection is always brighter than the threshold.
    \item \textit{Time-Bin Masking}: To simulate the macroscopic observational gaps caused by weather, telescope downtime, or seasonal visibility, we implement a temporal augmentation strategy adapted from the RAINBOW framework \citep{rainbow}. In this method, the total time span of the light curve is divided into uniform bins of $q$ days, where $q \in [0,5]$. We randomly select $p\%$ (where $p \in [0,50]$) of the bins that contain at least one true detection and then remove all photometric points within them by masking those temporal positions. These masked positions are subsequently ignored by the \texttt{AstraNet} encoder, effectively creating random $q$-day continuous gaps in the light curve. 
    \item \textit{Multi-Filter Random Window}: This is the integral temporal augmentation technique responsible for generating the multi-scale temporal views for \texttt{Astra-CLR}. We expand upon the single-filter random window strategy proposed in ASTROMER and adapt its mechanism to function effectively for multi-filter sequences (see Figure \ref{fig:random-win}). First, we iterate through each ZTF filter in the order ($g, r, i$) and extract a continuous window of detections from each, sampling the starting index of each window from a uniform distribution. The number of detections sampled per filter ($D$) is user-defined and explicitly determined by the inherent observational cadence of the ZTF survey, enforcing a strict hierarchy $D_r > D_g > D_i$. Finally, these individual filter windows are subsequently concatenated in the specific order ($g, r, i$) to construct a single unified \emph{view} of length $l$. The overall temporal size of this combined window is a hyperparameter, allowing us to generate both \textit{global} views (longer sequences) and \emph{local} views (shorter sequences). If any specific filter is entirely missing or contains fewer detections than its designated window size, we apply a padding mask to those positions. This ensures that the \texttt{AstraNet} encoder always receives a fixed-length input sequence while explicitly learning to robustly handle missing observational filters. 
\end{enumerate}
To visualize the combined effect of these augmentation strategies, Figure \ref{fig:original_lc} presents the original, unaugmented light curve of an active galactic nuclei (AGN). The corresponding multi-scale temporal views, comprising one global view and two distinct local views, are demonstrated in Figure \ref{fig:aug_lc}.

\subsection{Astra-CLR: Contrastive Learning Framework}
\label{astra-clr-framework}
\texttt{Astra-CLR} is a self-supervised representation learning framework built upon multi-crop contrastive paradigm. The framework processes and generates multiple asymmetric views of a single light curve, leveraging the data augmentation pipeline as detailed in Section \ref{sec:data-aug}.  These multi-scale temporal views comprise \emph{global views}  (GVs), which are longer sequences containing a combined total of over $200$ detections across all three filters ($g$, $r$, and $i$), and \emph{local views} (LVs), which are shorter sequences restricted to at most half the length of the GVs. Derived from the multi-filter random window strategy, a LV may either be a subsequence of the GV or a disjoint, non-overlapping sequence. All views originating from the same light curve collectively form a \textit{positive group}.

The architecture processes these asymmetric views through a pipeline of connected components. First, the complete input representation pipeline, $\mathbf{\tilde T}(\cdot)$, transforms the raw multivariate time-series by applying the previously detailed preprocessing and augmentation strategies (Section \ref{sec:data-prep} \& \ref{sec:data-aug}), and subsequently projecting the data into a dense vector space via \texttt{Astra Embeddings}. This embedding step combines the centered apparent magnitudes with temporal embeddings derived from a sinusoidal positional encoder. We then incorporate color information into the embeddings by mapping the ZTF filters through a non-linear Multi-Layer Perceptron (MLP). Second, the embedded sequence is passed to the \texttt{AstraNet} backbone, an encoder-only Transformer conceptually aligned with BERT \citep{bert} and TST \citep{zerveastransformer}, to extract a highly contextualized global representation of the irregularly sampled photometry. Third, we apply a non-linear MLP projection head to the backbone, mapping the output to a lower-dimensional latent space. Finally, the network is optimized using the Normalized Temperature-scaled Cross-Entropy (NT-Xent) loss function \citep{chen2017sampling,simclr}, which explicitly maximizes the latent similarity within a positive group. The mathematical and architectural formulations of these specific components are detailed in the subsequent subsections.

\subsubsection{Astra-CLR Embeddings}
\label{sec:astra-clr-emb}
\begin{figure}[t!]
	\centering \includegraphics[width=0.95\columnwidth, trim=1.cm 0.2cm 0.6cm 0.2cm, clip]{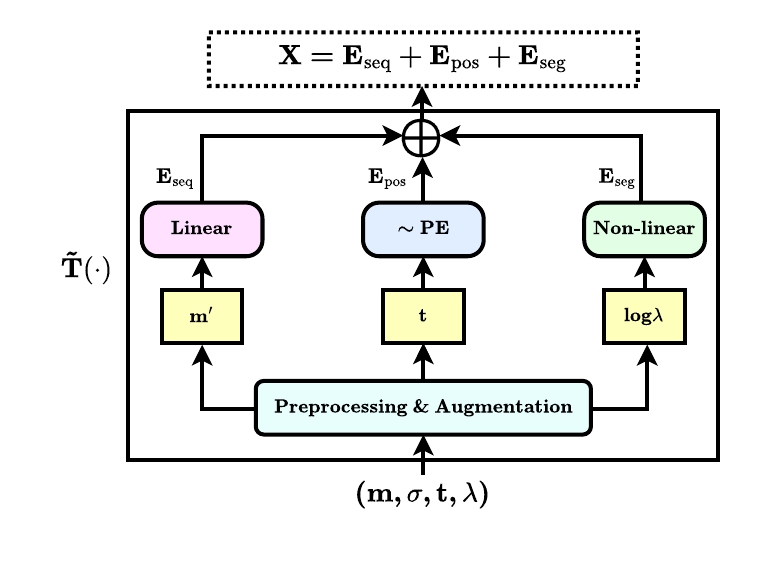}	
	\caption{Schematic overview of the input representation pipeline $\mathbf{\tilde{T}(\cdot)}$. A raw light curve $(m, \sigma, t, \lambda)$ is processed and augmented to yield features $(m', t, \log \lambda)$. These are independently projected into $d_{\text{model}}$-dimensional sequence ($\mathbf{E}_{\text{seq}}$), positional ($\mathbf{E}_{\text{pos}}$), and segment ($\mathbf{E}_{\text{seg}}$) embeddings. Their element-wise sum yields a final composite view ($\mathbf{X} = \mathbf{E}_{\text{seq}} + \mathbf{E}_{\text{pos}} + \mathbf{E}_{\text{seg}}$), which is subsequently passed to the \texttt{AstraNet} encoder.}
    \label{fig:inpu-emb}
\end{figure}
We transform the multivariate light curves into a composite embedding space before feeding them to the \texttt{AstraNet} encoder. As previously established in the preprocessing pipeline (see Section \ref{sec:data-prep}), we restructure each light curve as a unified sequence of detections grouped sequentially by the available \textsc{ZTF} filters (\textit{g, r, i}). Each observational point is parameterized by three primary components: the centered apparent magnitude ($m'$), the observation time in MJD ($t$), and the effective wavelength of the corresponding \textsc{ZTF} filters (i.e., $lg\lambda$).

The final input embedding matrix, $\mathbf{X} \in \mathbb{R}^{l \times d_{\text{model}}}$ (where $l$ is the variable sequence length of either a GV or LV, and $d_{\text{model}}$ is the output dimension of the \texttt{AstraNet} backbone), is the element-wise sum of three distinct vector representations:
\begin{enumerate}
    \item \textit{Sequence Embeddings ($\mathbf{E}_{\text{seq}}$):} The centered magnitudes ($m'$) are linearly mapped into a $d_{\text{model}}$-dimensional vector using a single fully connected layer without any non-linear activation.
    \item \textit{Positional Embeddings ($\mathbf{E}_{\text{pos}}$):} To capture the irregularity in the cadence of astronomical surveys and to effectively utilize the varying gaps introduced by our temporal masking during data augmentation, we employ a non-trainable, deterministic Positional Encoder (PE) to map the observational times into a $d_{\text{model}}$-dimensional vector. While standard NLP Transformers apply sinusoidal functions to discrete sequence indices \citep{attention, bert}, we use continuous observational times ($t$) in MJD to map them directly into sinusoidal embeddings, an approach also adapted by ASTROMER. In Equation \ref{eq:positional_encoding}, we define the PE using fixed sine and cosine function of different angular frequencies, where $j$ is the PE dimension and $\Omega$ is a frequency scaling constant used in calculating the angular frequencies.
    \begin{equation} 
    \label{eq:positional_encoding}
        PE_{(t, j)}= 
        \begin{cases} 
            \sin\left(t \cdot \Omega^{-j / d_{\text{model}}}\right) & \text{if } j \text{ is even} \\
            \cos\left(t \cdot \Omega^{(1-j) / d_{\text{model}}}\right) & \text{if } j \text{ is odd}
        \end{cases}.
    \end{equation}
    \item \textit{Segment Embeddings ($\mathbf{E}_{\text{seg}}$):} Since our framework utilize a multi-filter input sequence, we integrate $\lg\lambda$ as color and filter information into the sequences. Rather than employing standard categorical embeddings---which treat filters as discrete, mathematically unrelated classes---we explicitly pass the continuous $\lg\lambda$ scalars (i.e., $\lambda_{g}$$=4746.48\text{\AA}$, $\lambda_{r}$$=6366.38 \text{\AA}$, $\lambda_{i}$$=7829.03\text{\AA}$) for individual ZTF filters. We then project it into a $d_{\text{model}}$-dimensional vector space via a non-linear MLP utilizing a ReLU activation function. We deliberately utilize a non-linear function rather than a standard linear projection to ensure the network has the capacity to identify and model any complex, non-linear photometric relationships across the multi-filter sequence.
\end{enumerate}
To construct the final input representation for the network, we compute the element-wise sum of the individual sequence, position, and segment embeddings (i.e., $ \mathbf{X=E_{\text{seq}}+E_{\text{pos}}+E_{\text{seg}}}$) for individual views. This integrated embedding encapsulates all necessary photometry from the underlying light curve. It is then passed as an input to the \texttt{AstraNet} encoder to learn higher-order temporal dependencies. A visual schematic of this complete input embedding generation process is provided in Figure \ref{fig:inpu-emb}.

\subsubsection{Model Architecture: AstraNet}
\label{sec:astra-net}
\begin{figure}[t!]
	\centering \includegraphics[width=0.95\columnwidth, trim=0.9cm 0.9cm 1.2cm 0.9cm, clip]{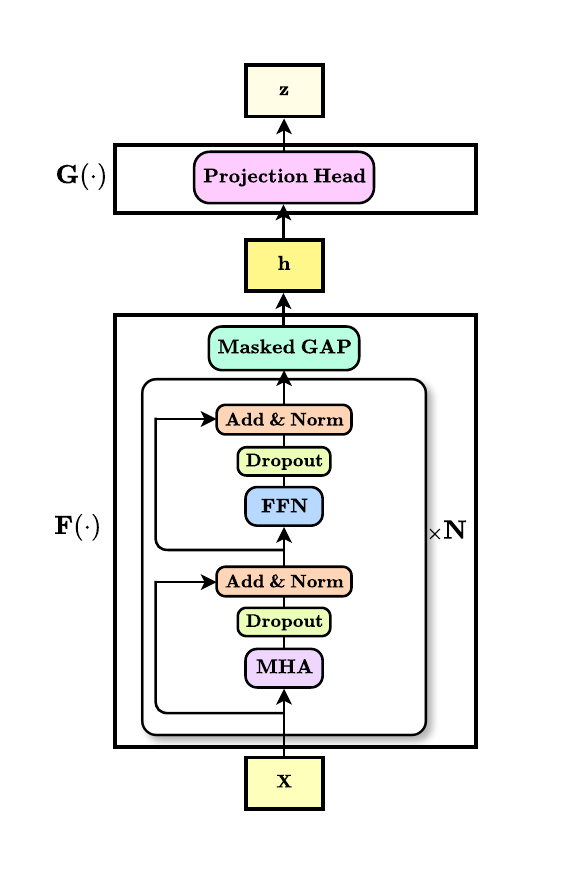}	
	\caption{Architectural schematic of \texttt{AstraNet}. The framework consists of a Transformer backbone, $\mathbf{F}(\cdot)$, and an MLP projection head, $\mathbf{G}(\cdot)$. The backbone processes the input view ($\mathbf{X}$) using a context-restricted Multi-Head Attention mechanism (Figure \ref{fig:astra-mha}), after which a masked 1D Global Average Pooling (GAP) layer condenses the contextualized sequence into a global representation vector ($\mathbf{h} \in \mathbb{R}^{d_{\text{model}}}$), explicitly filtering out invalid masked positions. Finally, $\mathbf{G}(\cdot)$ projects $\mathbf{h}$ into a lower-dimensional, $\ell_2$-normalized latent space ($\mathbf{z} \in \mathbb{R}^{d_{\text{proj}}}$) for contrastive optimization.}
    \label{fig:astra-net}
\end{figure}

\begin{figure}[t!]
	\centering \includegraphics[width=0.95\columnwidth, trim=0.6cm 1.1cm 0.6cm 1.2cm, clip]{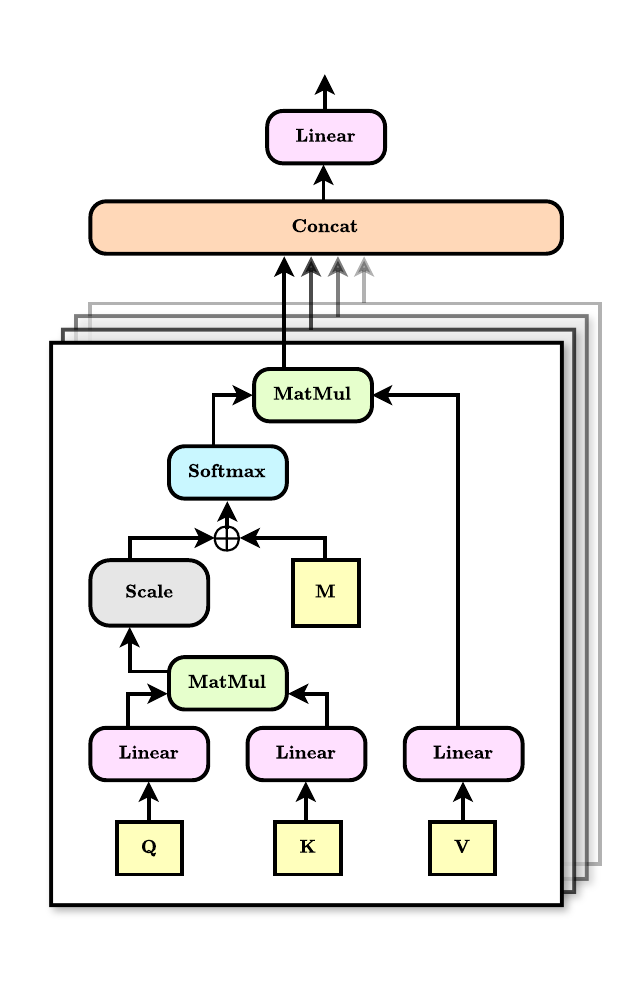}	
	\caption{Schematic of the context-restricted Multi-Head Attention (MHA) mechanism. This block introduces a critical modification to the standard Transformer: an additive penalty matrix $\mathbf{M}$. Adding $\mathbf{M}$ before the softmax ensures masked positions receive attention weights of exactly zero, so the model ignores artificially dropped or padded data as required by the data augmentation strategies.}
	\label{fig:astra-mha}
\end{figure}

\texttt{AstraNet} is the core representational architecture of our contrastive framework, as illustrated in Figure \ref{fig:astra-net}. It comprises a feature-extracting backbone, $\mathbf{F}(\cdot)$, followed by a projection head, $\mathbf{G}(\cdot)$. The backbone $\mathbf{F}(\cdot)$ incorporates an encoder-only Transformer---structurally adapted from the foundational NLP architecture proposed by \citet{attention}---coupled with a Global Average Pooling layer, similar to \citet{vit}, explicitly tailored to extract robust representations from astronomical time-series. The network processes the composite \texttt{Astra Embeddings} through a series of $N$ identical encoder layers. Each layer comprises two sub-layers---a context-restricted Multi-Head Attention (MHA) mechanism and a position-wise Feed-Forward Network (FFN). The output of each sub-layer is summed to its input through a residual connection, followed by layer normalization. To facilitate these residual connections across all sub-layers in the model, including the embedding layers, we use a fixed output dimension $d_{\text{model}}$ in our final configuration.\\
\newline\textbf{Attention Mechanism and Masking Strategy}\\
Because the multi-scale temporal views inherently lack certain photometric filters and contain intentionally masked temporal positions from our augmentation pipeline, a rigorous masking mechanism is required during the self-attention phase. We implement a specific masking protocol within the scaled dot-product attention to ensure the model strictly ignores these missing or artificially masked observations. As illustrated in Figure \ref{fig:astra-mha}, the attention mechanism operates on queries $\mathbf{Q}\in \mathbb{R}^{l \times d_{\text{k}}}$, keys $\mathbf{K}\in \mathbb{R}^{l \times d_{\text{k}}}$, and values $\mathbf{V}\in \mathbb{R}^{l \times d_{\text{v}}}$, where $l$ denotes the sequence length and $d_v$ and $d_k$ represent the dimensions within each attention head. 

We define a binary mask vector $\mathbf{m}\in \{0,1\}^{1 \times l}$, where $\mathbf{m}_{j}$=$1$ if the temporal position $j$ is invalid and should be ignored, and $\mathbf{m}_{j}$=$0$ if it is a valid, unmasked position. During the attention computation, the masked vector $\mathbf{m}$ is broadcasted across the query dimension to a 2D additive penalty matrix $\mathbf{M}\in \mathbb{R}^{l\times l}$. The masked attention output matrix is then calculated as:
\begin{equation}
    \text{Attention}(\mathbf{Q}, \mathbf{K}, \mathbf{V}) = \text{softmax}\left(\frac{\mathbf{Q}\mathbf{K}^T}{\sqrt{d_k}} + \mathbf{M}\right)\mathbf{V},
\end{equation}
where the elements of the additive penalty matrix $\mathbf{M}_{ij}$ is defined by,
\begin{equation}
    \mathbf{M}_{ij} =
    \begin{cases}
    -\infty & \text{if } \mathbf{m}_{j} = 1 \\
    0 & \text{if } \mathbf{m}_{j} = 0
    \end{cases}.
\end{equation}
The infinite negative penalty drives the subsequent softmax activation for masked positions to zero, completely nullifying their contribution to the output sequence.\\
\newline\textbf{Pooling and Projection Head}\\
The output of the final Transformer encoder layer is a contextualized sequence matrix, $\mathbf{y} \in \mathbb{R}^{l \times d_{\text{model}}}$. We aggregate this sequence into a unified, fixed-length representation by applying a masked 1D Global Average Pooling (GAP) operation across the temporal dimension $l$. We utilize the binary mask vector $\mathbf{m}$ to exclude invalid positions across the temporal index while applying the GAP operation. This combined process of the Transformer encoder and the GAP layer defines our representational backbone, $\mathbf{F}(\cdot)$, which maps the input embeddings, $\mathbf{X}$, to a rich, global representation vector $\mathbf{h} = \mathbf{F}(\mathbf{X}) \in \mathbb{R}^{d_{\text{model}}}$.

We adapt our contrastive learning framework following the conventions established by SimCLR, where the global representation vector, $\mathbf{h}$, is not directly used for computing the contrastive loss. Instead, it is passed through a non-linear MLP projection head, $\mathbf{G}(\cdot)$, to improve the representational quality of the preceding layer embeddings (i.e., $\mathbf{h}$). The projection head consists of a linear dense layer with ReLU non-linear activation, followed by a final linear projection layer mapping to the $d_{\text{proj}}$-dimensional latent space. We apply $\ell_2$ normalization to the output of the projection head, $\mathbf{z} = \mathbf{G}(\mathbf{h})$, such that $\|\mathbf{z}\|_2 = 1$. We subsequently use the final normalized vector, $\mathbf{z}$, to compute the NT-Xent contrastive loss.\\
\newline\textbf{Regularization and Training Stability}\\
We explicitly introduce regularization within the encoder layers to prevent the model from overfitting. We apply dropout layers $(drop\_rate=0.1)$ on the outputs of both the MHA and FFN sub-layers, prior to their respective residual connections and layer normalization. Furthermore, we employ gradient clipping before updating the model's weights to ensure numerical stability during the optimization step. Because deep neural architectures are highly susceptible to exploding gradients---which can cause disproportionately large and unstable weight updates---we restricted the global norm of the gradients to a maximum of $1.0$. This constraint bounds the maximum step size taken by the optimizer, preventing catastrophic divergence and ensuring training stability.

\subsubsection{Contrastive Loss Function: NT-Xent}
\label{sec:nt-xent-loss}
For the \texttt{Astra-CLR} framework, we adopt the \textbf{N}ormalized \textbf{T}emperature-scaled \textbf{Cross}-\textbf{Ent}ropy (NT-Xent) loss function as introduced in SimCLR. However, we generalize this formulation to handle our multi-scale temporal view strategy by extending it beyond the standard two-view pairwise comparisons to simultaneously handle multiple asymmetric views (i.e., a combination of GVs and LVs) for each light curve.

Let $B$ denote the number of original light curves in a given training batch (i.e., the original batch size). After processing each sequence through the augmentation pipeline, we transform each light curve into $n$ distinct views, resulting in an augmented batch size of $B'=nB$. In our final architectural configuration, we used $n=3$ for pre-training. Following the projection head $\mathbf{G}(\cdot)$, we obtain a set of $\ell2$-normalized representation vectors, $\mathbf{z}$, which are unit vectors used to optimize the NT-Xent loss.
Consequently, the cosine similarity between any two unit vectors, $\mathbf{z}_j$ and $\mathbf{z}_k$, strictly simplifies to their dot product as follows:
\begin{equation}
\label{eq:sim_matrix}
    \text{sim}(\mathbf{z}_j, \mathbf{z}_k) = \mathbf{z}_j^\top \mathbf{z}_k.
\end{equation}
In our multi-scale temporal framework, all $n$ views originating from the same light curve constitute a \textit{positive group} (i.e., $\{\mathbf{z}_1, \mathbf{z}_2, \dots, \mathbf{z}_n\}$). Within each \textit{positive group}, we generate bidirectional positive pairs by comparing every view against all other views in the same group, explicitly excluding self-similarity. This yields $n(n-1)$ positive pairs per light curve. For any given projection ($\mathbf{z}_j$) within the augmented batch, we treat the remaining $(B'-n)$ views---which originate from different light curves---as negative samples. We then formulate a contrastive predictive task by optimizing an objective that maximizes the latent similarity among the views within each positive group, while simultaneously minimizing the similarity between each projection  ($\mathbf{z}_j$) and its $(B'-n)$ negative samples,
\begin{equation}
\label{pos_pair_loss}
    \ell_{j,k} = - \log \frac{\exp\left(\text{sim}(\mathbf{z}_j, \mathbf{z}_k) / \tau\right)}{\sum_{w=1}^{B'} \mathbbm{1}_{[w \neq j]} \exp\left(\text{sim}(\mathbf{z}_j, \mathbf{z}_w) / \tau\right)},
\end{equation}
where $\tau > 0$ denotes the temperature hyperparameter that scales the distribution of the logits, and $\mathbbm{1}_{[w \neq j]}$ is an indicator function evaluating to $1$ if $w \neq j$ and $0$ otherwise, explicitly excluding self-similarity from the denominator.

Since our framework utilizes multiple asymmetric views, we compute the loss across all positive pairs within each \textit{positive group}. The final objective function minimized by \texttt{Astra-CLR} is the mean loss across all positive pairs in the augmented batch of size $B'$:
\begin{equation}
\label{total_loss}
    \mathcal{L}_{\text{total}} = \frac{1}{|P|} \sum_{j=1}^{B'} \sum_{k \in P(j)} \ell_{j,k},
\end{equation}
where $P(j)$ denotes the set of indices of all positive counterparts of $\mathbf{z}_j$ in a \textit{positive group}, and $|P|$ is the total number of positive pairs evaluated in the training batch (i.e., $|P| = n(n-1)B$).

\begin{figure*}[t!]
	\centering 
    \includegraphics[width=\textwidth, trim=1.cm 1.cm 1.cm 0.cm, clip]{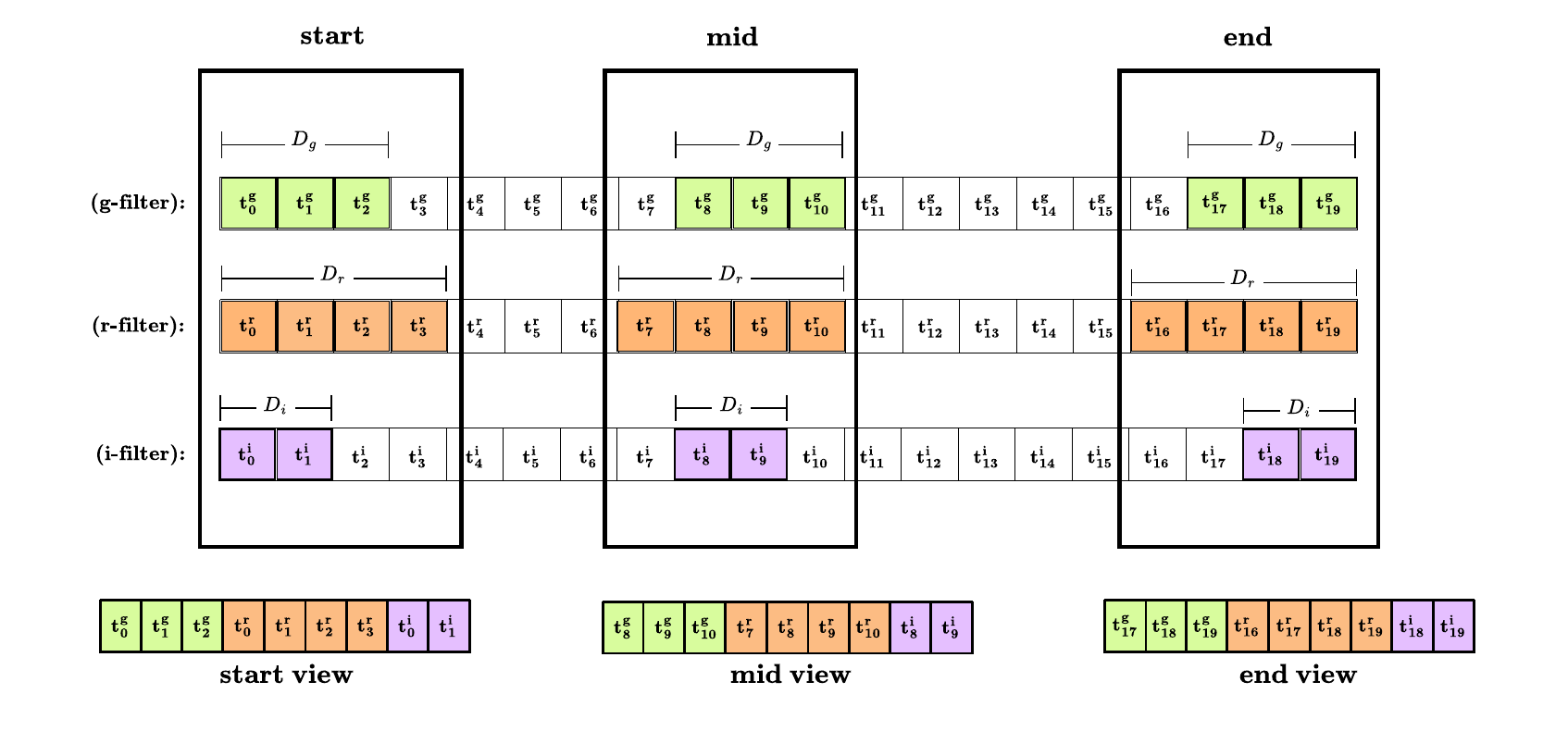}
	\caption{Schematic diagram of the temporal views employed in the late fusion architecture. For illustrative purposes, we assume the original light curve sequence contains 20 consecutive detections per photometric filters - ($g$, $r$, $i$). Following Equation \ref{eq:three-window}, the algorithm extracts a deterministic, continuous temporal window from each filter. These segments are subsequently concatenated along the temporal axis in the strict order of $[g, r, i]$ to construct three fixed-length views ($l=9$): \textbf{start}, \textbf{mid}, and \textbf{end}. Each view is processed independently through the \texttt{AstraNet} encoder, and their respective outputs are concatenated to yield the final fused global representation vector, $\mathbf{h}_{\text{fused}} \in \mathbb{R}^{3d_{\text{model}}}$.}
	\label{fig:three_window}
\end{figure*}

\subsubsection{Optimizer and Learning Rate Scheduling}
\label{optandlr}
We used the Adam optimizer \citep{adam} with $\beta_1$=$0.9$, $\beta_2$=$0.98$, and $\epsilon$=$10^{-9}$. We varied the learning rate ($lr$) over the course of training using the Noam scheduling strategy \citep{attention}:
\begin{equation}
    \label{noam}
    lr = d_{\text{model}}^{-0.5} \cdot \min \left(step^{-0.5}, step \cdot warmup\_steps^{-1.5} \right).
\end{equation}
This scheduler increases the learning rate linearly for the first $warmup\_steps$ training steps, and then decreases proportionally to the inverse square root of the $step$ number. We used a $warmup\_steps$$=1000$ for pre-training \texttt{Astra-CLR}.

\subsection{ Downstream Feature Extraction via Multi-View Late Fusion}
\label{late_fusion}
To maximize contextual information extracted from the variable-length light curves, we introduce a deterministic multi-view late fusion strategy utilized across all downstream evaluations (illustrated in Figure \ref{fig:three_window}). While late fusion techniques are widely established in computer vision \citep{snoek2005early,karpathy2014large}, their application to astronomical time series provides a novel mechanism to capture both short-term and long-term periodic behaviors within a single forward pass, ensuring a strictly consistent baseline for comparing pre-trained and fine-tuned representations.

For a given multi-filter light curve $\mathbf{x}$, we define $L_g$, $L_r$, and $L_i$ as the total number of available detections in the $g$, $r$, and $i$ filters, respectively. For each ZTF filter, $f \in \{g, r, i\}$, we define a target window size $D_f$, such that the total sequence length for any generated view is $l = D_g + D_r + D_i$, maintaining a strict hierarchy $D_r > D_g > D_i$ (e.g., $D_g=300$, $D_r=350$, $D_i=50$). We extract three distinct temporal views ($v \in \{\text{start, mid, end}\}$) by first independently slicing each filter sequence $\mathbf{x}_f$ and subsequently combining them to a unified temporal view $\mathbf{x}_{f, v}$ as defined below:
\begin{equation}
\label{eq:three-window}
    \mathbf{x}_{f, v} = 
    \begin{cases} 
    \mathbf{x}_f[0 : D_f] & \text{if } v = \text{start} \\
    \mathbf{x}_f\left[c - \lfloor \frac{D_f}{2} \rfloor : c + \lfloor \frac{D_f}{2} \rfloor\right] & \text{if } v = \text{mid} \\
    \mathbf{x}_f[L_f - D_f : L_f] & \text{if } v = \text{end}
    \end{cases},
\end{equation}
where the midpoint of $\mathbf{x}_{f}$ is defined as $c = \lfloor \frac{L_f}{2} \rfloor$. Conversely, if a specific filter sequence is shorter than its target length ($L_f < D_f$), we dynamically right-pad the sequence with zeros to reach $D_f$, and fill the corresponding mask vector, $\mathbf{m}_f$, with ones for these padded steps. Once the filter-specific slices are extracted, they are concatenated along the temporal axis to construct the unified multi-filter view:
\begin{equation}
\mathbf{x}_v = \mathbf{x}_{g, v} \parallel \mathbf{x}_{r, v} \parallel \mathbf{x}_{i, v},
\end{equation}
where $\parallel$ denotes the concatenation operator. As formalized in our self-attention mechanism (Section \ref{sec:astra-net}), the binary mask vector strictly ensures that the \texttt{AstraNet} encoder ignores these padded masked observations during temporal aggregation.

Each of the three views is independently passed through the \texttt{AstraNet} encoder, $\mathbf{F}(\cdot)$, yielding three $d_{\text{model}}$-dimensional embeddings. These intermediate embeddings are then fused via concatenation along the feature axis to construct the unified, global representation vector:
\begin{equation}
\mathbf{h}_{\text{fused}} = \mathbf{F}(\mathbf{x}_{\text{start}}) \parallel \mathbf{F}(\mathbf{x}_{\text{mid}}) \parallel \mathbf{F}(\mathbf{x}_{\text{end}}).
\end{equation}
This results in a comprehensive $(3 d_{\text{model}})$-dimensional latent vector, $\mathbf{h}_{\text{fused}}$.

\section{Training Details and Implementation}
\label{sec:training-config}

This section details the empirical optimization and computational deployment of the \texttt{Astra-CLR} framework. We begin by outlining a comprehensive hyperparameter search (Section \ref{sec:hyper-params-opt}), exploring architectural variations, optimization dynamics, and data-centric constraints unique to astronomical time-series. Building on these empirical findings, we subsequently define the finalized pre-training configuration and specify the distributed hardware infrastructure utilized for the final large-scale pre-training (Section \ref{sec:final-pretraining-setup}).

\begin{table*}[t!]
    \centering
    \caption{Downstream task evaluation on the $120$K class-balanced dataset. We report the accuracy, micro F1-score, and macro F1-score (all in \%) for both the pre-trained and fine-tuned \texttt{Astra-CLR} models. Evaluations were conducted using linear probing and weighted $k$-NN protocols. As shown, the label-efficient fine-tuning strategy consistently yields superior performance across all metrics compared to the frozen pre-trained baseline.}
    \label{tab:downstream_results}
    \vspace{5pt}
    \renewcommand{\arraystretch}{1.3} 
    \begin{tabular*}{\textwidth}{@{\extracolsep{\fill}}lcccc}
        \toprule
        \textbf{Model Configuration} & \textbf{Evaluation Protocol} & \textbf{Accuracy} & \textbf{Micro F1-Score} & \textbf{Macro F1-Score} \\
        \midrule
        \multirow{2}{*}{\texttt{Astra-CLR} (Pre-trained)} 
        & linear    & $70.34\pm0.14$ & $70.34\pm0.14$ & $70.27\pm0.14$ \\
        & $k$-NN        & $62.32\pm0.20$ & $62.32\pm0.20$ & $62.17\pm0.20$ \\
        \midrule
        \multirow{2}{*}{\texttt{Astra-CLR} (Fine-tuned)} 
        & linear              & $77.23\pm0.13$ & $77.23\pm0.13$ & $77.26\pm0.13$ \\
        & $k$-NN    & $76.42\pm0.13$ & $76.42\pm0.13$ & $76.42\pm0.13$ \\
        \bottomrule
    \end{tabular*}
\end{table*}

\subsection{Hyperparameter Optimization}
\label{sec:hyper-params-opt}
To establish an optimal architectural and training configuration with minimal computational overhead, we accelerated our comprehensive hyperparameter search by downscaling the pre-training dataset to a representative subset of approximately 21 000 light curves. We then evaluated the representational quality of each configuration using a downstream linear probing protocol, which we discuss later in Section \ref{sec:downstream_tasks_pt}. To ensure statistical robustness, the performance metrics for each configuration were stabilized via $100$ bootstrap resampling iterations.
\newline \newline\textbf{Architectural Search}: We began by optimizing the structural hyperparameters of the \texttt{AstraNet} backbone, followed by its projection head, to establish our base encoder. We varied the network depth across $N\in\{4,6,8\}$ layers, revealing that a shallower architecture of $N=4$ encoder blocks significantly outperforms deeper variants, likely avoiding overfitting on the sparse time-series data. We evaluated latent embedding dimensions, $d_{\text{model}}\in\{256,512\}$, to assess the impact of significantly scaling the model's overall parameter counts, and observed comparable downstream performance across both configurations. For the continuous-time positional encoding, we adjusted the frequency scaling constant, $\Omega$, to a lower-frequency value of 1 000, and observed that the standard Transformer adaptation of 10 000 as proposed by \citet{attention}, drastically outperformed it. Finally, we varied the output dimension of the projection head across $d_{\text{proj}}\in\{128,256,512\}$. We concluded that keeping $d_{\text{proj}}$ to the encoder output dimension ($d_{\text{model}}$) or exactly half of its size ($d_{\text{model}}/2$) resulted in comparable downstream results. In addition to these structural hyperparameter searches, we conducted a comprehensive Ablation Study (see Section \ref{sec:ablation_studies}) for an in-depth analysis of core architectural choices. This includes a comparative analysis of symmetric versus asymmetric multi-scale temporal view configurations, an investigation into an alternative broad-spectrum positional encoding formulation, and an empirical evaluation of the necessity of multi-filter segment embeddings.
\newline \newline\textbf{Optimization and Loss Dynamics}: We trained the models using the Adam optimizer coupled with the learning rate scheduler detailed in Section \ref{optandlr}. When tuning the duration of the learning rate warmup phase (i.e., $warmup\_steps$) across \{1 000, 2 000, 4 000, 8 000\} steps, we observed that shorter warmup periods resulted in significantly faster convergence and superior final representations. Additionally, we tuned the temperature scalar of the NT-Xent loss function ($\tau\in\{0.1,1.0,10.0\}$). We found that $\tau=1.0$ provided the optimal scaling of the similarity logits, effectively regulating the softmax distribution in Equation \ref{pos_pair_loss}.
\newline \newline\textbf{Data-Centric Dynamics}: Finally, our hyperparameter search implicated critical insights regarding data dimensions that contrast sharply with the behavior of canonical contrastive self-supervised models. First, standard contrastive frameworks in computer vision---such as SimCLR, MoCo, and CLIP---typically require massive batch sizes to ensure an adequate diversity of negative samples. For instance, in the astronomical domain, AstroCLIP utilizes a batch size of $1024$ for its multimodal contrastive alignment, crucial for the InfoNCE loss, which is inherently upper-bounded by the number of negative samples. Similarly, foundational models pre-trained on highly diverse benchmark datasets like IG-1B \mbox{\citep{instagram}}, ImageNet \mbox{\citep{imagenet, imagenet-1k, imagenet-21k}}, and JFT-300M \mbox{\citep{jft-300m}} utilize standard large batches ranging from $1024$ to $4096$ for pre-training to capture the dataset's distribution. Our pre-training dataset is distributed across twelve broad variable star classes, and moderate batch sizes ranging from $128$ to $640$ provide a sufficient statistical representation of the underlying class distribution within a single training step. Increasing the batch size for \texttt{Astra-CLR} highlighted no measurable improvement in embedding quality. However, employing larger batches helps optimize hardware throughput and accelerate training times. Second, we evaluated the impact of scaling the multi-filter sequence lengths across all generated views. We tested maximum GV lengths of $l \in \{300, 500, 700, 900\}$ detections, with the corresponding LVs proportionally scaled to exactly half the GV length. We observed a strong positive correlation between the input sequence length and downstream representation quality. This demonstrates that the \texttt{Astra-CLR} framework performs optimally when provided with longer input sequences, enabling the multi-scale architecture to effectively capture and leverage long-term temporal dependencies through the \textit{local-to-global} mapping strategy.

\subsection{Final Pre-training Configuration and Hardware Details}
\label{sec:final-pretraining-setup}
The final \texttt{Astra-CLR} model contains $\sim13$ million trainable parameters. The \texttt{AstraNet} backbone consists of $N=4$ encoder layers with an output dimension\footnote{In the standard positional encoder \citep{attention}, each pair of embedding dimensions corresponds to a distinct sinusoid. Utilizing $d_{\text{model}}=512$ provides a sufficiently large embedding space to capture a high-resolution spectrum of frequencies, resulting in a richer, multi-scale positional representation.}of $d_{\text{model}}=512$ and a projection head with a dimension, $d_{\text{proj}}=256$. We integrate sequence, positional, and segment embeddings to form the final inputs for \texttt{Astra-CLR}. During data loading, the augmentation pipeline constructs three contrasting views per light curve: one global view (GV) truncated at $700$ detections, and two distinct local views (LVs) restricted to $\sim350$ detections.

The final pre-training run was accelerated via a distributed TensorFlow \citep{tensorflow} architecture across eight instances of NVIDIA A100 GPUs ($80$ GB VRAM) on a single machine, efficiently scaling to a global batch size of $624$ ($78$ per replica) to maximize throughput. The network was pre-trained for exactly $200$ epochs, which required $\sim10.3$ days of total training time.

\section{Results}
\label{sec:results}
In this section, we systematically evaluate the efficacy, robustness, and discriminative power of the representations learned by the \texttt{Astra-CLR} framework. We begin by assessing the inherent quality of the frozen pre-trained latent space using standard downstream evaluation protocols---specifically linear probing and weighted $k$-NN classification---to establish a strong self-supervised baseline. Building upon this foundation, we demonstrate the framework's remarkable label efficiency. We introduce a partial top-layer fine-tuning strategy that utilizes $2\%$ of the available ground-truth labels to analyze how minimal supervised intervention can dramatically refine the local geometry and global decision boundaries of the embeddings. Finally, we present a comprehensive comparative analysis between the pre-trained and fine-tuned architectures. This quantitative evaluation is supported by high-resolution hierarchical UMAP visualizations---provided in \ref{sec:umap}---which qualitatively confirm the distinct, taxonomic clustering achieved by the model.

\subsection{Downstream Evaluation of Pre-Trained Representations}
\label{sec:downstream_tasks_pt}
To evaluate the discriminative power of the learned representations, we utilize two standard downstream task evaluation protocols: linear probing and weighted $k$-NN classification. For these evaluations, we curated a class-balanced dataset comprising $120,000$ training light curves and $24,000$ validation light curves, uniformly distributed across the $12$ variable star classes. 

To extract the static feature vectors for these baseline evaluations, we employ the deterministic multi-view late fusion strategy detailed in Section \ref{late_fusion}. We extract the three distinct temporal views (start, mid, and end) from each light curve, pass them independently through the frozen \texttt{AstraNet} backbone, $\mathbf{F}(\cdot)$, and concatenate them to yield the unified $3 d_{\text{model}}$-dimensional representation vector, $\mathbf{h}_{\text{fused}}$. Notably, the pre-trained representations are inherently robust; even when evaluating using a single, randomly sampled temporal window, the downstream performance remained highly competitive. However, utilizing the concatenated three-view representations ensures maximum temporal coverage and establishes a strict comparative baseline for our subsequent fine-tuning evaluations.

For the linear probing protocol, a Multinomial Logistic Regression classifier is trained directly on these frozen embeddings. This protocol explicitly tests the linear separability of the latent space; the capacity of the self-supervised model to disentangle the underlying class distributions without any task-specific weight updates.

Complementing linear probing, we employ a weighted $k$-NN classifier on the same frozen representations. Because $k$-NN requires no parametric training, it assigns labels based strictly on latent proximity. Following the evaluation strategy established by the DINO framework \citep{dino}, we apply $\ell_2$ normalization to the feature vectors and utilize cosine similarity (Equation \ref{eq:sim_matrix}) to compute neighbor distances. This protocol directly assesses the metric quality and topological structure of the latent space, testing its inherent capability to cluster distinct taxonomic classes. 

The performance metrics, overall accuracy, micro F1-score, and macro F1-score, for both evaluation protocols are summarized in Table \ref{tab:downstream_results}. Notably, achieving $>70\%$ linear probing accuracy across $12$ complex variable star classes, without the model ever being exposed to ground-truth labels during pre-training, strongly indicates that the contrastive loss function successfully captured the fundamental physics and distinct morphological features inherent to the light curves.

\begin{figure*}[t!]
	\centering 
    \includegraphics[width=\textwidth]{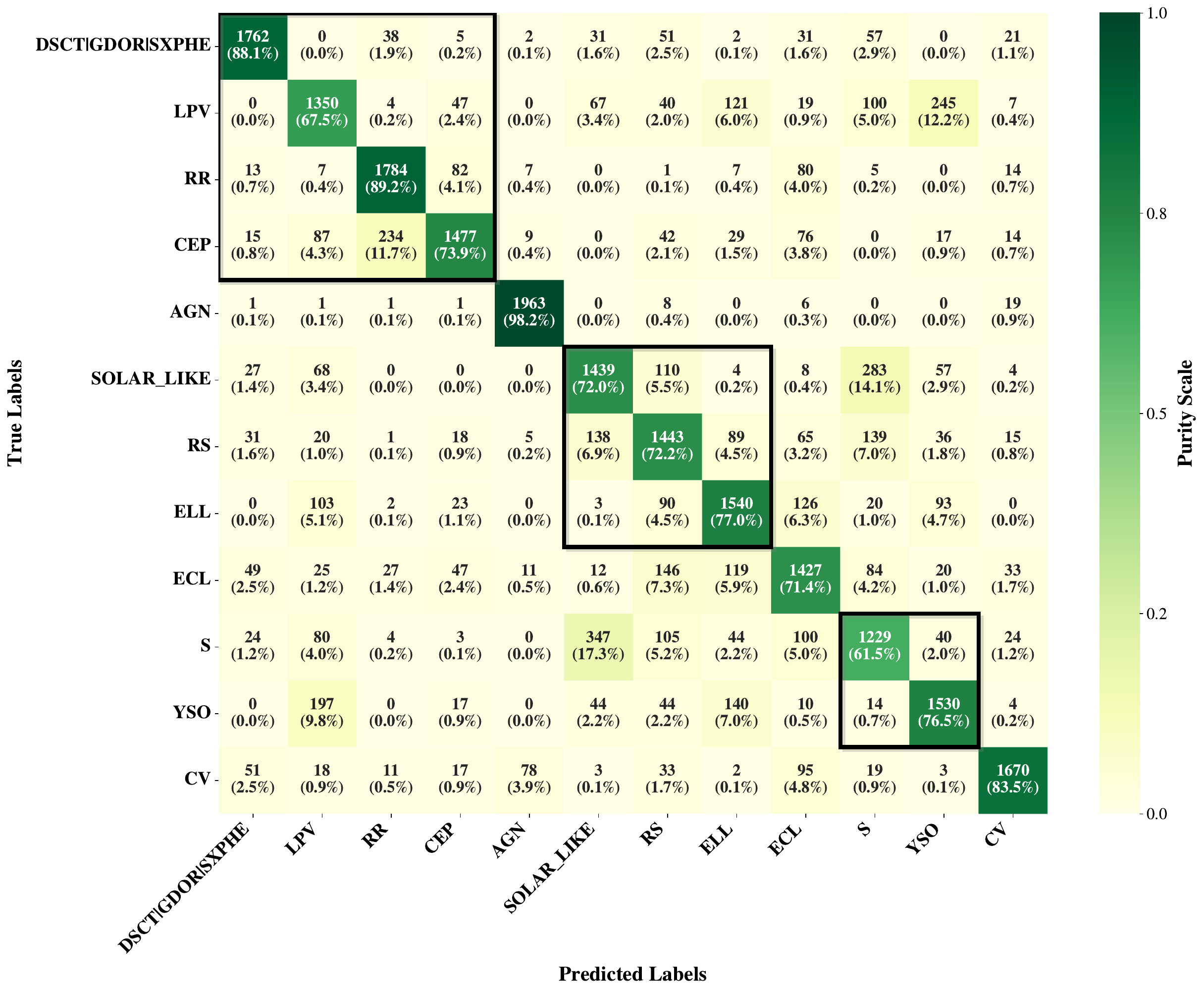}
	\caption{Linear probing evaluation of the fine-tuned \texttt{Astra-CLR} representation on the $24$K balanced validation dataset. The matrix displays both absolute classification counts and row-wise predictive purity (in \%) across the $12$ variability classes. The highlighted rectangular blocks indicate the broader astronomical groupings, corresponding to the categorization previously outlined in Table \ref{tab:variable_dist}.}
	\label{fig:confusion_matrix}
\end{figure*}

\subsection{Label-Efficient Partial Fine-Tuning}
\label{pft}
While the frozen pre-trained representations establish a robust self-supervised baseline, real-world astronomical surveys require models that can adapt to specific classification tasks with minimal annotation effort. To evaluate the label efficiency of \texttt{Astra-CLR}, we introduce a highly constrained supervised fine-tuning strategy utilizing $2\%$ of the original pre-training dataset ($\sim$ 42 000 light curves). By coupling a deterministic multi-view late fusion technique with partial top-layer unfreezing, we demonstrate that, with minimal supervised intervention, we can dramatically refine the model's representational boundaries. The following subsections (Sections \ref{ft_data} \& \ref{tft}) detail our task-specific data curation and architectural adaptations. Subsequently, we present the resulting performance enhancements in Section \ref{sec:downstream_tasks_ft}.
\subsubsection{Task-Specific Data Curation}
\label{ft_data}
To ensure high-quality inputs for supervised refinement, we first applied centered apparent magnitude scaling to the light curves, as formulated in Equation \ref{eq:standardization_step}. Furthermore, to maintain a strict signal-to-noise baseline during training, we restricted the candidate pool to relatively bright objects. Specifically, we filtered for light curves possessing a weighted mean apparent magnitude (Equation \ref{eq:weighted_mean_sub}) of $\bar{m}_w < 18$ in at least one of their observational filter. Following these preprocessing constraints, we sampled $2\%$ of the original pre-training dataset, yielding a highly curated set of 41 827 labeled light curves for the fine-tuning phase.

\begin{figure*}[t!]
    \centering
    \includegraphics[width=\textwidth]
    {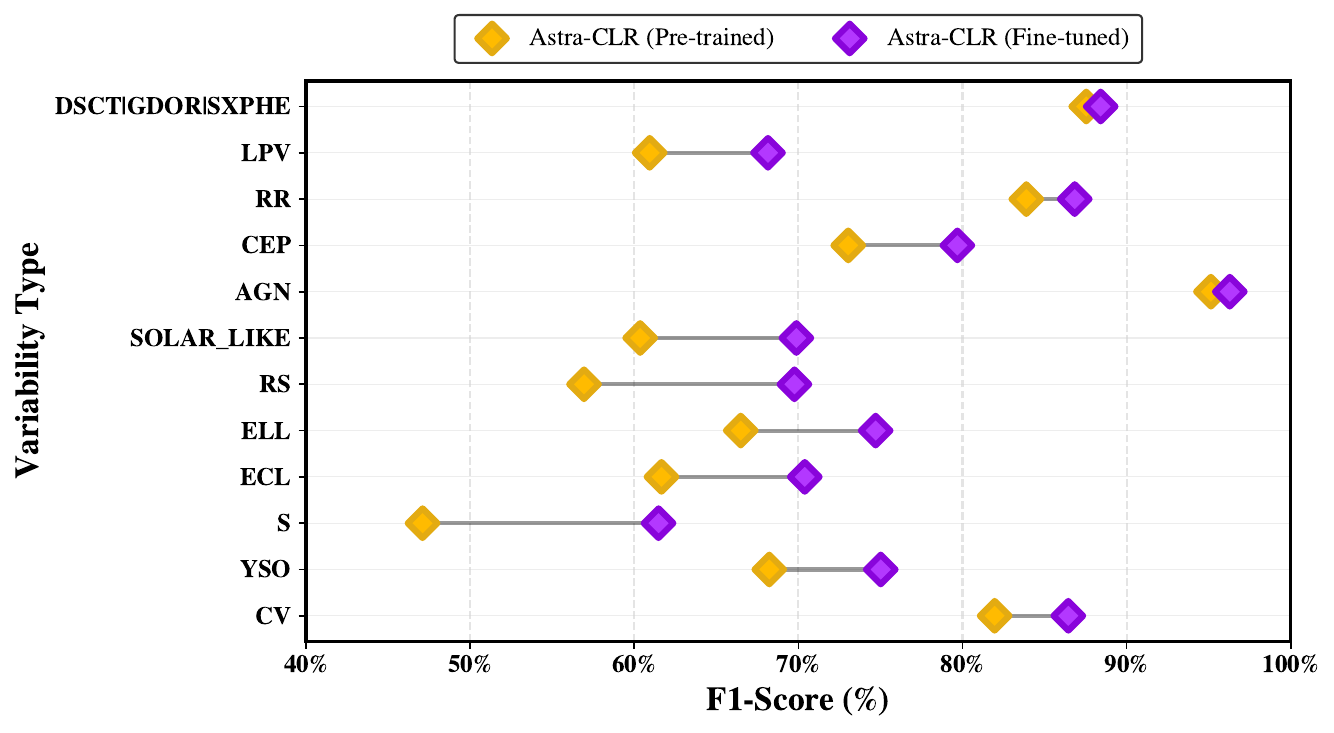}
    \caption{Comparative linear probing performance of the \texttt{Astra-CLR} framework. The visualization illustrates the F1-score progression from the frozen pre-trained representations to the fine-tuned embeddings across all $12$ variable star classes, highlighting the consistent performance gains achieved through label-efficient fine-tuning.}
    \label{fig:f1-scores}
\end{figure*}

\subsubsection{Top-Layer Fine-Tuning Protocol}
\label{tft}
To configure the supervised network, we initialized the \texttt{AstraNet} encoder, $\mathbf{F}(\cdot)$, with our contrastive pre-trained weights. We implemented a partial unfreezing strategy: the majority of the lower network layers were strictly frozen, while only the final two Transformer encoder blocks were left trainable. This approach ensures the retention of the robust, generalizable low-level features learned during self-supervised pre-training, while explicitly allowing the deeper, high-level representations to adapt to the specific classification task. 

During the forward pass, we employed the multi-view late fusion strategy (detailed in Section \ref{late_fusion}). The start, mid, and end temporal views of each curated light curve were processed independently through the partially unfrozen encoder and concatenated to form the unified $3d_{\text{model}}$-dimensional representation, $\mathbf{h}_{\text{fused}}$. Following this fusion step, we appended a dropout layer ($drop\_rate=0.2$) for regularization. Subsequently, we attach a dense classification head mapping to the target classes. The unfrozen encoder blocks and the newly initialized classification head were optimized using the Adam optimizer with a sparse categorical cross-entropy loss objective. We executed the training on a single NVIDIA A100 GPU (80 GB VRAM) for $100$ epochs, incorporating an early stopping patience of $10$ epochs to prevent over-fitting.

\subsection{Downstream Performance of Fine-Tuned Representations}
\label{sec:downstream_tasks_ft}
To quantify the impact of our label-efficient partial fine-tuning, we re-evaluated the \texttt{Astra-CLR} embeddings using the identical linear probing and weighted $k$-NN protocols. During inference, we strictly froze the fine-tuned architecture. We extracted the unified $3 d_{\text{model}}$-dimensional representation, $\mathbf{h}_{\text{fused}}$, for each light curve---maintaining the exact multi-view late fusion extraction protocol used for the pre-trained baseline---and fed these static embeddings directly to the classifiers. 

The fine-tuning phase yielded substantial performance gains: overall linear probing accuracy improved from $\sim70\%$ to $\sim 77\%$, while $k$-NN accuracy experienced a dramatic boost from $\sim 62\%$ to $\sim 76\%$ (see Table \ref{tab:downstream_results}). This $14\%$ jump in $k$-NN performance mathematically confirms that the top-layer refinement successfully tightened the local neighborhood geometry of the latent space, pulling instances of the same class closer together while pushing distinct classes further apart. 

To provide a granular view of this improved classification dynamics, Figure \ref{fig:confusion_matrix} presents the purity matrix of the fine-tuned embeddings evaluated via linear probing. The strong diagonal dominance and minimal off-diagonal confusion demonstrate high predictive purity across the $12$ variable star classes, clearly delineating the variability groups. 

Furthermore, to explicitly quantify the representational shift, Figure \ref{fig:f1-scores} visualizes the per-class F1-score progression from the frozen pre-trained state to the fine-tuned state. The plot illustrates that while supervised fine-tuning provides critical performance gain---particularly for morphologically complex variability classes---the pre-trained self-supervised baseline was already highly competitive. Ultimately, this quantitative success is qualitatively mirrored by the tightened topological clustering observed in the fine-tuned hierarchical UMAP projections (see \ref{sec:umap}).

\section{Discussion}
\label{sec:discussion}

Our evaluation highlights that domain-specific data augmentation is not just beneficial, but fundamentally necessary to learn robust representations. In our framework, augmentation serves as the primary engine for generating multi-scale temporal views, which proved to be a critical component of downstream performance (Table \ref{tab:multiview}). Specifically, the representational quality improved significantly when transitioning from two symmetric global views to an asymmetric configuration (1 GV and 1 LV). Performance scaled even further with a three-view strategy (1 GV and 2 LVs), demonstrating that the model learns richer structural invariants when forced to execute a \emph{local-to-global} mapping strategy. Furthermore, the integration of segment embeddings proved highly effective, playing a critical role in capturing complex, non-linear photometric relationships across the multi-filter sequence.

Interestingly, our experiments revealed a nuanced dynamic regarding contrastive training configurations. Standard contrastive frameworks typically prioritize massive batch sizes to maximize the number of negative samples. However, we observed that for datasets with smaller class distributions (in our case, $12$ broader variability types), the benefits of aggressively scaling the batch size diminish. Instead, expanding the temporal sequence length yielded a much more prominent performance boost. This suggests that for complex astronomical time series, maximizing the context of long-term dependencies available within individual sequences is more critical than maximizing inter-sample contrast across a batch.

Finally, the implementation of our \emph{multi-view late fusion} architecture during the fine-tuning phase successfully addressed a fundamental challenge in time-domain astronomy: the extreme disparity in observational cadences. Because the \texttt{AstraNet} encoder requires fixed-length input sequences, processing real-world light curves, which in our dataset range from $5$ to over 4 000 individual photometric detections per filter, presents a significant bottleneck. By extracting and independently processing start, mid, and end temporal views, we effectively sampled across the entire observational baseline. This ensured that long-term dependencies were captured without requiring aggressive data truncation. When coupled with our label-efficient fine-tuning strategy---updating only the top two layers using $2\%$ of the labeled data---this fusion mechanism successfully refined the high-level semantic features. This refinement is quantitatively validated by the substantial improvements in our weighted $k$-NN metrics, confirming that the framework effectively tightened the local neighborhood geometry of the latent space.

As detailed by \citet{Rimoldini_GaiaVariability_2023}, the $12$ overarching Gaia variability types used in our downstream evaluation encompass over $35$ nested variable subtypes. Furthermore, the Gaia taxonomy includes broad (``generic'') classes, such as \texttt{SOLAR\_LIKE} and \texttt{S} (Short-timescale). During the compilation of the Gaia classification training set from external literature, these generic labels were assigned primarily to variable sources that lacked a more highly resolved physical classification. As a consequence of this class mapping strategy, the \texttt{SOLAR\_LIKE} class broadly aggregates magnetically active stars exhibiting generic spot or flare dynamics. Similarly, the \texttt{S} class groups sources defined primarily by the rapid speed of their photometric variations rather than by specific physical emission mechanisms. Since the \texttt{S} class aggregates fast-varying objects from diverse astrophysical backgrounds that lack a specific dedicated classification, it effectively acts as a broad morphological class of variability. This underlying heterogeneity in the training data perfectly explains its centralized, bridging position within our UMAP projection (see \ref{sec:umap}). Notably, our partial fine-tuning strategy was particularly effective at disentangling these broad classes from other morphologically similar variability types, significantly improving their downstream classification metrics.

During the fine-tuning phase, we uniformly sampled 4 000 labeled light curves for the majority of variability types, with the exceptions of Cepheids (CEP, $967$ samples) and Cataclysmic Variables (CV, $860$ samples) due to their limited counts in the preprocessed catalog (see Table \ref{tab:variable_dist}). This deliberate sampling distribution allowed us to systematically isolate which specific variability types benefit most from supervised refinement. Our evaluation revealed a nuanced dynamic between the chosen evaluation metric and the topological changes induced by fine-tuning.

Through linear probing, we observed that highly distinct classes, such as \texttt{AGN} and \texttt{DSCT|GDOR|SXPHE}, already achieve robust linear separability strictly through self-supervised pre-training. Interestingly, supervised partial fine-tuning slightly stagnated the linear probing performance for these already well-isolated classes. This occurs because the supervised cross-entropy objective in the fine-tuning phase attempts to overfit the global decision boundaries to rare, noisy outliers. For instance, artificially stretching the linear hyperplane of the \texttt{DSCT|GDOR|SXPHE} class to accommodate training outliers accidentally encroaches upon the neighboring latent spaces of \texttt{RS}, \texttt{RR}, or \texttt{S} (Short-timescale)---variability types, thereby increasing linear misclassification rates for an otherwise perfectly separable cluster.

However, this boundary distortion is strictly a limitation of the linear probing evaluation rather than a true stagnation of the learned representations. When evaluated via weighted $k$-NN classification---which relies on local neighborhood geometry rather than rigid global hyperplanes---performance saw a significant boost across all classes, with every variability type improving on their F1-score. This sharp contrast between two metrics confirms that while partial top-layer fine-tuning may warp global decision boundaries to fit outliers, it consistently succeeds in pulling intra-class samples into tighter, denser local clusters without compromising the underlying representational quality.

\section{Summary and conclusions}
\label{sec:conclusion}
In this work, we presented \texttt{Astra-CLR}, a robust self-supervised representation learning framework designed explicitly for multi-filter astronomical time-series data. Driven by a novel multi-scale contrastive instance discrimination approach tailored for photometry, the model successfully constructs a highly discriminable latent space for variable stars, pre-trained entirely on an unlabeled dataset of $\sim 2.1$ million Zubercal light curves. 

Central to this success is the introduction of domain-specific data augmentation techniques, which serve as the primary engine for generating multi-scale temporal views (global and local views) and integrating cross-filter segment embeddings. Furthermore, to accommodate different cadences while satisfying the fixed-length constraints of our architecture, we introduced a novel multi-view late fusion mechanism. By seamlessly integrating this extraction strategy, our framework effectively captures both short-term varying trends and long-term periodic dependencies within an underlying light curve.

Comprehensive downstream evaluations rigorously validate the efficacy of this architecture. While the frozen pre-trained representations inherently established a highly competitive linear baseline (achieving $>70\%$ accuracy), evaluating the framework via weighted $k$-NN classification perfectly illustrates the impact of our label-efficient fine-tuning. By applying a partial top-layer fine-tuning strategy, we significantly refined the high-level semantic features, particularly by disentangling broad ``generic'' classes (i.e., \texttt{SOLAR\_LIKE} and \texttt{S}) from other complex variability types. This minimal supervised intervention resulted in a substantial $14\%$ performance jump in weighted $k$-NN accuracy (from $\sim 62\%$ to $\sim 76\%$). Because weighted $k$-NN classification is strictly sensitive to local data density, this drastic improvement quantitatively confirms the tightening of the latent space's local neighborhood geometry---a result beautifully mirrored in our qualitative hierarchical UMAP projections.

\paragraph{Future Work}
While \texttt{Astra-CLR} establishes a highly effective foundation for self-supervised light curve analysis, several promising avenues remain for future development. Architecturally, we aim to design a dynamic positional encoder that jointly embeds time in MJD ($t$) and photometric color information ($\lg \lambda$) to better capture the underlying temporal information in a multi-filter light curve. At the survey level, we plan to investigate cross-survey transfer learning---evaluating how well representations pre-trained on one catalog (e.g., Zubercal) generalize to disparate observational domains (e.g., ASAS-SN, TESS, or OGLE) without retraining. This adaptability will be crucial as we scale the framework to process the unprecedented data influx anticipated from next-generation facilities such as the \textit{Vera C. Rubin Observatory} and the \textit{Nancy Grace Roman Space Telescope}. Furthermore, we intend to leverage our tightly clustered fine-tuned latent space for real-time anomaly detection, enabling the automated flagging of rare, out-of-distribution transients in live alert streams. Finally, we are actively expanding the broader \texttt{ASTRA} ecosystem; a novel, non-contrastive representation learning framework for astronomical time series is currently under development and slated for upcoming release.

\begin{table*}[t!]
\centering
\caption{Downstream evaluation of multi-scale temporal view generation strategies on the $120$K class-balanced dataset. Performance is measured using accuracy, micro F1-score, and macro F1-score (all reported in \%). The results demonstrate that the asymmetric three-view strategy significantly outperforms both two-view configurations: the dual global view (2 GVs) setup and the mixed global-local view (1 GV + 1 LV) setup.}
\label{tab:multiview}
\vspace{5pt}
\renewcommand{\arraystretch}{1.3} 
\begin{tabular*}{\textwidth}{@{\extracolsep{\fill}}llcccc}
\toprule
\textbf{Views} & \textbf{View Configuration} & \textbf{Evaluation Protocol} & \textbf{Accuracy} & \textbf{Micro F1-Score} & \textbf{Macro F1-Score}\\
\midrule

\multirow{2}{*}{2} & \multirow{2}{*}{2 GVs ($l=700$)} 
& linear & $57.48\pm0.17$ & $57.48\pm0.17$ & $57.36\pm0.17$ \\
& & $k$-NN & $55.28\pm0.26$ & $55.28\pm0.26$ & $55.07\pm0.26$ \\
\midrule

\multirow{2}{*}{2} & \multirow{2}{*}{1 GV ($l=700$) + 1 LV ($l=350$)} 
& linear & $62.15\pm0.20$ & $62.15\pm0.20$ & $62.05\pm0.20$ \\
& & $k$-NN & $57.43\pm0.23$ & $57.43\pm0.23$ & $57.25\pm0.23$ \\
\midrule

\multirow{2}{*}{\textbf{3 (Base)}} & \multirow{2}{*}{\textbf{1 GV ($l=700$) + 2 LVs ($l=350$)}} 
& linear & \bm{$70.34\pm0.14$} & \bm{$70.34\pm0.14$} & \bm{$70.27\pm0.14$} \\
& & $k$-NN & \bm{$62.32\pm0.20$} & \bm{$62.32\pm0.20$} & \bm{$62.17\pm0.20$} \\

\bottomrule
\end{tabular*}
\end{table*}

\section{Software Availability}
\label{sec:software}

To facilitate reproducibility and community adoption, all resources associated with this work are made publicly available:
\begin{itemize}
    \item \textbf{Source Code:} The complete codebase for pre-training, fine-tuning, and downstream evaluation is hosted on GitHub under the \texttt{astra} project\footnote{\url{https://github.com/TorshaMajumder/astra}}.
    \item \textbf{Inference Package:} For streamlined deployment, we provide \texttt{astra-infer}, a lightweight Python package available in a separate repository\footnote{\url{https://github.com/snad-space/astra-infer}}.
    \item \textbf{Model Weights:} All pre-trained and fine-tuned weights, alongside optimized ONNX files, are accessible via the Hugging Face Hub\footnote{\url{https://huggingface.co/ashrot/astra-clr-base}}.
\end{itemize}

\section*{Acknowledgments}
We thank Maria Pruzhinskaya for valuable discussions, thoughtful feedback, and helpful suggestions throughout this work.

We gratefully acknowledge the NVIDIA Academic Grant Program for providing the primary high-performance hardware and software infrastructure critical to this research. Specifically, all core code development and distributed pre-training of the \texttt{Astra-CLR} framework were executed on an NVIDIA Brev cloud instance equipped with eight NVIDIA A100 (80 GB VRAM) GPUs and supported by the integrated NVIDIA CUDA toolkit.
Additionally, we acknowledge the support of Cloud computing and AI technology provider \texttt{Cloud.ru}\footnote{\url{https://cloud.ru/}} for providing the computational resource to conduct all hyper-parameter testing and optimization. 
This work used Bridges-2~\citep{sanielevici2021bridges} at Pittsburgh Supercomputing Center for the HATS catalog preparation through allocation PHY210095 from the Advanced Cyberinfrastructure Coordination Ecosystem: Services \& Support (ACCESS) program, which is supported by National Science Foundation grants \#2138259, \#2138286, \#2138307, \#2137603, and \#2138296. 
Furthermore, we extend our gratitude to the Cosmostatistics Initiative\footnote{\url{https://cosmostatistics-initiative.org/}} (COIN) and the Laboratoire de Physique de Clermont Auvergne (LPCA) for providing the server infrastructure utilized for downstream metric evaluation and the generation of all manuscript visualizations.
Support was provided by Schmidt Sciences, LLC. for K.~Malanchev.


\appendix
\section{Ablation Studies}
\label{sec:ablation_studies}
We empirically validated the architectural and methodological design choices of the \texttt{Astra-CLR} framework by conducting ablation studies on the multi-scale temporal view configuration (focusing on the number of views and its length), the application of a scaled Positional Encoder ($PE^s$) versus the canonical $PE$ (Equation \ref{eq:positional_encoding}), and the influence of segment embeddings. Unless otherwise specified, the ablated models were trained on the full $2.1$ million pre-training dataset under identical optimization conditions used for the final pre-training setup. We performed downstream evaluation of the ablated models using the identical linear probing and weighted $k$-NN classification, utilizing the same $120$K balanced dataset as discussed in Sections \ref{sec:downstream_tasks_pt} and \ref{sec:downstream_tasks_ft}.
\newline \newline\textbf{Impact of Multi-scale Temporal View Configuration}\newline
A core contribution of the \texttt{Astra-CLR} framework is the adaptation of the ``multi-crop'' strategy\footnote{Each view in \texttt{Astra-CLR} was randomly generated via a distinctly parameterized data augmentation pipeline as discussed in Section \ref{sec:data-aug}.} for astronomical time-series. To justify this architectural choice, we compared standard two-view contrastive models against our finalized three-view approach.
\begin{enumerate}
    \item \textit{Two-View (Symmetric)}: The model was pre-trained using two global views (GVs) of equal length per light curve ($l=700$), mirroring the traditional SimCLR approach.
    \item \textit{Two-View (Asymmetric)}: The model was pre-trained using one GV ($l=700$) and one local view (LV)($l=350$).
    \item \textit{Three-View Multi-Scale (Final Model)}: The model was pre-trained using one GV ($l=700$) and two independent LVs ($l=350$).
\end{enumerate}
As detailed in Table \ref{tab:multiview}, our results demonstrate that the three-view multi-scale strategy significantly outperforms both two-view architectures. By forcing the network to simultaneously match multiple augmented LVs to the broader global sequence, the three-view approach compels the model to identify any strong underlying correlations among the augmented light curves. Furthermore, this multi-scale configuration introduces beneficial variance into the contrastive views. Consequently, the network is compelled to learn deeper, multi-scale physical correlations in a \textit{positive group}, yielding a highly robust time-invariant representation space.
\newline \newline\textbf{Influence of Positional Encoding Scaling}
\newline Given the irregular cadence of the photometric data, we investigated the PE formulation through a broader frequency range of [10 mins, 100 000 days]. In this experimental setup, we isolated the positional embeddings---keeping all other network parameters strictly identical to the 3-view architecture during pre-training. We introduced a scaling factor ($\rho$) to modulate the angular frequencies and expanded the temporal dimensionality to $d_{\text{model}}$ \citep{astromer}. Unlike the standard $PE$, which pairs sine and cosine functions at identical frequencies, our scaled formulation strictly operates at distinct frequencies:
\begin{equation} 
\label{eq:scaled_positional_encoding}
    PE^s_{(t, j)} = 
    \begin{cases} 
        \sin\left(t\cdot \rho \cdot \Omega^{-j / d_{\text{model}}}\right) & \text{if } j \text{ is even} \\
        \cos\left(t\cdot \rho \cdot \Omega^{-j / d_{\text{model}}}\right) & \text{if } j \text{ is odd}
    \end{cases}.
\end{equation}
We evaluated the scaled $PE^s$ performance on the 120K balanced dataset, and it achieved a linear probing accuracy of $23.58\pm0.24$ \% and a $k$-NN accuracy of $18.91\pm0.23$ \%. Ultimately, this approach did not surpass the performance of the unscaled continuous-time $PE$ formulation (linear: $70.34\pm0.14$ \%, $k$-NN: $62.32\pm0.20$ \%).\\
\newline\textbf{The Necessity of Segment Embeddings}\\
The \texttt{Astra-CLR} architecture processes multi-filter light curves as a unified, ordered, and concatenated sequence of \textit{g, r, and i}-filter detections. To prove the necessity of explicitly passing color information, we conducted a structural ablation in which the segment embeddings were entirely removed. Because this architectural evaluation was conducted during the initial hyperparameter testing phase, the ablated configuration was pre-trained using $21$K light curves rather than the full $2.1$ million pre-training dataset. In this ablated configuration, the input embeddings were generated using only the sequence and positional embeddings, completely ignoring the non-linear MLP segment embeddings. This ablation resulted in a quantifiable degradation in linear probing accuracy, empirically validating the necessity of explicitly encoding filter information in \texttt{Astra Embeddings}.

\begin{figure*}[p] 
    \centering
    \includegraphics[width=\textwidth, trim=0.cm 0.0cm 0.cm 0.0cm, clip]{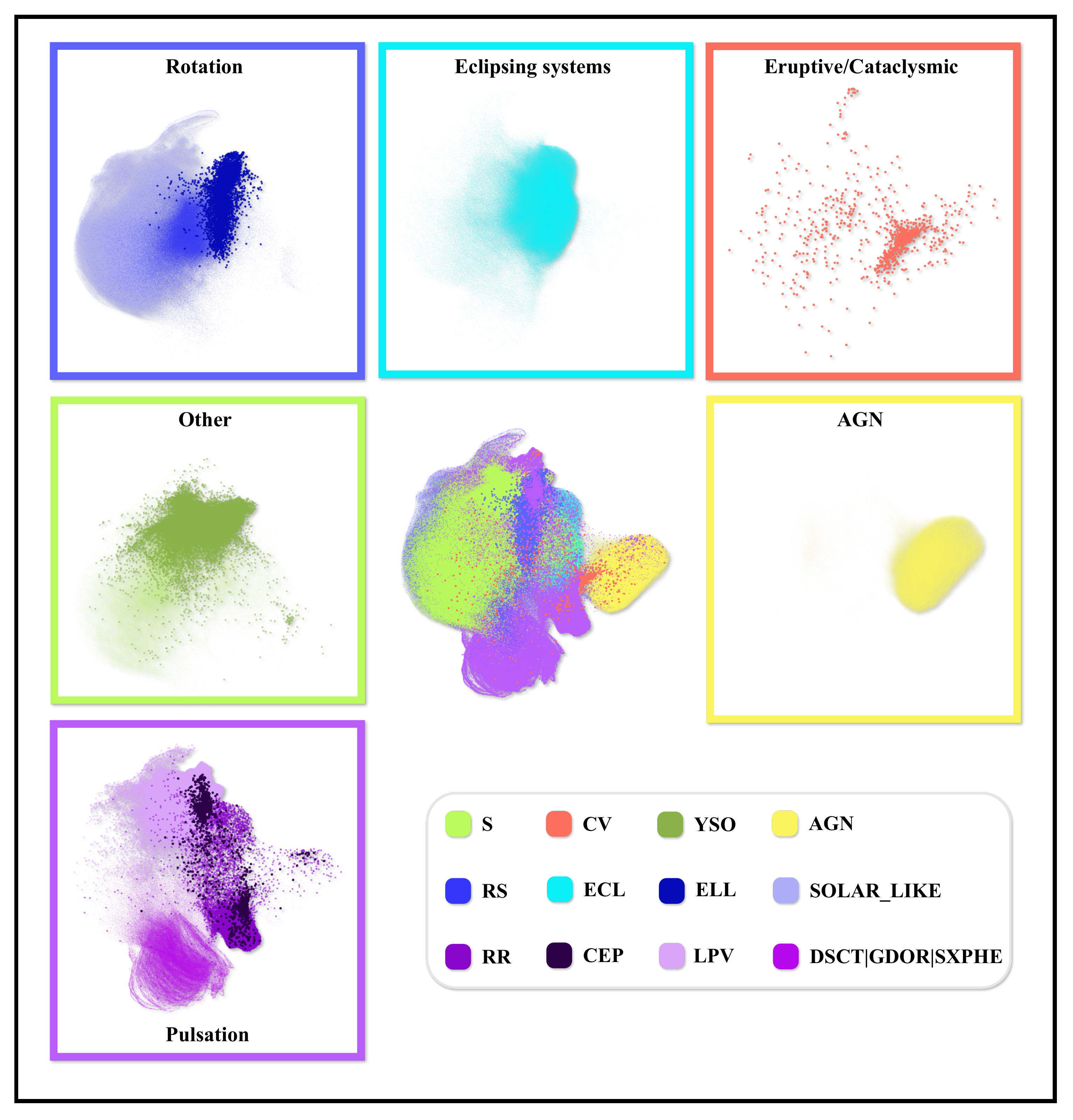}
    \caption{Hierarchical UMAP projection of the fine-tuned \texttt{Astra-CLR} latent space, encompassing $\sim 2.1$ million light curves. \textbf{Center:} The global latent space distribution colored by the six overarching variability groups outlined in Table \ref{tab:variable_dist}. This demonstrates the framework's capacity to form tightly clustered, highly discriminable astronomical classes following label-efficient partial fine-tuning, consistent with its strong $k$-NN classification performance ($>76\%$). \textbf{Surrounding Inlets:} Localized, magnified projections of each individual group. The colored borders of the inlet panels correspond directly to the broader color mapping in the central hub. Within each inlet, constituent sub-classes are visualized using high-contrast gradients of the parent color, as detailed in the single shared legend at the bottom. The strong topological separation of these sub-classes---such as the distinct isolation of AGN from other variability types---highlights the fine-grained representational robustness achieved by the fine-tuned \texttt{Astra-CLR} model.}
    \label{fig:umap_ft}
\end{figure*}

\section{Qualitative Analysis of the Fine-tuned Astra-CLR Latent Space}
\label{sec:umap}
To perform a qualitative assessment of class separability and clustering capability within the learned latent space, we project the high-dimensional \texttt{Astra-CLR} representations into a 2D plane using Uniform Manifold Approximation and Projection (UMAP) \mbox{\citep{umap}}. The resulting topological structure of the label-efficient, partially fine-tuned embeddings is illustrated in Figure \ref{fig:umap_ft}.

\clearpage

\bibliographystyle{elsarticle-harv} 
\bibliography{example}






\end{document}